# Cometary Chemistry and the Origin of Icy Solar System Bodies, the view after Rosetta

Kathrin Altwegg[1], Hans Balsiger[1] and Stephen A. Fuselier[2,3]

[1]Space Research and Planetary Sciences, University of Bern, CH-3012 Bern, Switzerland

[2] Space Science Directorate, Space Science and Engineering Division, Southwest Research Institute, San Antonio, TX 78238, USA

[3]University of Texas at San Antonio, San Antonio, TX 78249, USA

Email: altwegg@space.unibe.ch; hans.balsiger@space.unibe.ch; sfuselier@swri.edu

ORCID Altwegg: 0000-0002-2677-8238

ORCID Fuselier: 0000-0003-4101-7901

## Abstract

In situ research on cometary chemistry began when measurements from the Giotto mission at comet 1P/Halley revealed the presence of complex organics in the coma. New telescopes and space missions have provided detailed remote and in situ measurements of the composition of cometary volatiles. Recently, the Rosetta mission to comet 67P/Churyumov-Gerasimenko more than doubled the number of parent species and the number of isotopic ratios known in comets. 39 of the 66 parent species have also been detected in pre- and proto-stellar clouds, making the similarities of cometary ices with pre-stellar material very intriguing. Most isotopic ratios are non-solar. The variations in D/H in water between different comets indicate a large range in the protoplanetary disk where comets formed. All of these results point to a non-homogeneized protoplanetary disk where comets received their material. This diverse origin is in contrast to the Sun, who received its material from the bulk of the collapsing cloud. Detection of $N_2$ and Ar in the coma of 67P and the presence of very volatile $S_2$ point to low formation temperature of 20-30 K of the comet. The xenon isotopic ratios measured in 67P can explain the long standing question about the origin of the terrestrial atmospheric xenon. While we can exclude comets as being the source of the bulk terrestrial water due to their generally high D/H, the noble gases in the Earth’ atmosphere are compatible with a cometary delivery. This means that the amount of organics delivered by comets may be highly significant.

# 1 Introduction

Questions about the origin of the matter that makes up our solar system, the formation of planets, and finally the emergence of life on Earth are some of the big topics in astrochemistry. By now we probably understand the cycle from the diffuse interstellar material to dark molecular clouds and star forming regions and finally to protoplanetary disks and planetary systems like our own (see fig. 1). The basis of the observable material in the interstellar medium originates from remnants of stars, supernovae and neutron star mergers, and gas and dust. This material still preserves the signature of its origin through different isotopic ratios of the elements. Already at this stage, chemistry sets in, although densities and temperatures are very low. Once the densities become higher in dark clouds, chemistry becomes more important, especially chemistry on dust grain surfaces. Thanks to observatories like the Atacama Large Millimeter/submillimeter Array (ALMA) we are now at the stage where ongoing formation of protostars and young protoplanetary disks are regularly observed in detail. These observations include dust and some of their chemical molecular inventory. ALMA has enabled the study of many complex molecules in clouds and disks, revealing an extremely rich chemical inventory. At the other end of stellar system evolution, we are making progress in exploring the make-up of planets and moons in our own solar system. The questions, however, remain: how much of the material we see in our solar system has been inherited from the pre-solar stage, how much chemistry was ongoing during the protosolar and protoplanetary disk phase, and how well is the material preserved in some of the most primitive bodies in our solar system since their formation. Comets are an important link between the planets and moons of today and pre-solar and protoplanetary chemistry in clouds and disks.

In 1950 Fred Whipple (Whipple, 1950) postulated the “dirty snowball”, the idea that comets consist of ice and dust. At the same time, Oort (1950) postulated the idea that the comet reservoir is an isotropic cloud around, but very far from the Sun. This led to the notion that comets represent some of the most primitive material in our solar system, conserving their material from 4.6 billion years ago. This notion is still correct, although we now know more about cometary reservoirs, not just Oort cloud, but also Kuiper belt / scattered disk and we no longer look at comets as dirty snowballs but rather icy dirtballs.

Research on cometary chemistry started more than 100 years ago with the detection of the optical line of CN in the tail of comet 1P/Halley in 1910. By 1986, before the encounters of the Russian VEGA spacecraft and the European Giotto mission with comet 1P/Halley, several other species were known to exist, mostly radicals and ions like $CO^+$, CN, CH, $CO_2^+$ , $N_2^+$ and NH. The water molecule as such had not yet been detected, but the presence of water was inferred from OH. The appearance of comet Halley in 1985/86 triggered a huge effort to observe coma composition (refractories and gas) from several space missions as well as from ground-based observatories. As expected, water was detected by several instruments on Giotto (e.g. Krankowsky, 1986) as well as for the first time from the ground (Mumma, 1986). However, it became very clear that the chemistry in comets is much more complex than anticipated. Measurements of the Picca sensor on Giotto revealed the existence of molecules up to at least mass 100 da (Korth et al., 1986). One of the most intriguing measurements was the D/H ratio in water. Two separate measurements from the Giotto mission, one from the Ion Mass Spectrometer (IMS, Balsiger et al., 1995), and the other one from the Neutral Mass Spectrometer (NMS, Eberhardt et al., 1995) gave almost identical results. Their combined value of $\sim(3.1 \pm 0.4)\times10^{-4}$ is twice the terrestrial value. This result alone showed clearly that comets must have formed at low temperatures and must have conserved their original composition.

In the subsequent decades, cometary chemistry was pushed forward by a) bigger telescopes and space missions (e.g. IRAM30m and Herschel) which opened up new frequency bands with enhanced sensitivity and b) the arrival of two bright comets, C/1996 B2 (Hyakutake) and C/1995 O1 (Hale-Bopp). These advances increased the number of detected parent molecules to 27 (e.g. Bockelée-Morvan et al., 2000; Biver et al., 2002). With the large number of molecules, comparison between comets and comet families became feasible. Comets were classified according to their chemical diversity (Mumma & Charnley, 2011; Dello Russo et al., 2016). Meanwhile, advances in astrochemical modelling and in radio observations of the interstellar material in molecular clouds and star forming regions revealed similar molecular abundances for several complex molecules to those of cometary coma, hinting at a similar origin (Bockelée-Morvan et al., 2000; Schöier et al., 2002).

While remote sensing enables comparison of different comets, the observations are biased by the presence / absence of molecular transition lines in the frequency bands available and biased by the brightness of the comet. Therefore, many more observations exist for Oort cloud comets (OCC's) than for Jupiter family comets (JFC's), because OCC's are generally larger and brighter. For example, IR observations have determined the chemical composition of 30 comets, but only 9 of these are JFC's (dello Russo et al., 2016). In addition, most observations are done close to perihelion. Except for a few very bright comets, it is not possible to observe the coma composition during an extended part of the cometary orbit.

Several space missions encountered comets after Giotto, but all of them were flybys where the observation time is very short (Stardust, Deep Impact, EPOXI). None of them had any in situ capabilities to explore the cometary coma. With the exception of Stardust which brought back refractory material from comet 81P/Wild 2, all of them provided only a snapshot of the cometary coma at the time of encounter. Rosetta was the first mission to accompany a comet for a large part of its journey around the Sun, measuring the release of gas and dust during the different phases. As a result of this unique mission concept, all major and many minor species including their isotopologues were measured as a function of heliocentric distance and as a function of latitude/longitude of the comet nucleus.

Rosetta was able to characterize many features of 67P during its orbit around the Sun. It more than doubled the known cometary parent species and added many more isotopic ratios to the cometary inventory. Furthermore, these measurements were made with higher precision than previous, limited observations, which makes it possible to draw more stringent conclusions about the origin of cometary ices and the formation of these icy bodies. That means we have currently quite a number of comets where the main and major minor species in the coma are characterized and we have one comet where we know the composition in great detail. For 67P, we know all major species, minor species down to relative abundances of $10^{-7}$ and isotopes, sublimation characteristics, heterogeneities in the coma and variations along the orbit of the comet.

In this paper we focus on the volatile inventory of the cometary ices, as volatiles are most affected by chemistry. By analyzing the coma of comets and comparing the findings from Rosetta and previous comet observations with observations of the ISM, of pre-solar nebula, star forming regions and protoplanetary disks, we try to follow the chemical and physical processes which took place from the ISM to comets and finally to the Earth and life. This review is a post-Rosetta update on earlier reviews on cometary chemistry ( Mumma & Charnley, 2011; Cochran et al., 2015; Bockelée-Morvan et al., 2015).

In the next chapter we give a short overview on the Rosetta mission as many results in the following chapters originate from Rosetta/ROSINA measurements.

# 2 The Rosetta mission

Recognizing the importance of comets for the history of our solar system, the European Space Agency ESA started planning a big cometary mission already before the launch of Giotto in 1985. As part of the Horizon 2000 program the cometary mission Rosetta was selected for flight in 1995 (Glassmeier et al.,2007). The mission goals were

- Global characterization of the nucleus, determination of dynamic properties, surface morphology and composition
- Chemical, mineralogical and isotopic compositions of volatiles and refractories in a cometary nucleus
- Physical properties and interrelation of volatiles and refractories in a cometary nucleus:
- Study the development of cometary activity and the processes in the surface layer of the nucleus and in the inner coma (dust-gas interaction)
- Origin of comets, relationship between cometary and interstellar material. Implications for the origin of the solar system

The payload consisted of 11 instruments and a lander, which itself had a payload of 9 miniature instruments. Several of the instruments were designed to explore the composition of the nucleus and coma, among them specifically the Rosetta Orbiter Sensor for Ion and Neutral Analysis (ROSINA) (Balsiger et al., 2007). The scientific goals of ROSINA consisted of determining the elemental, isotopic, and molecular composition of the comet's atmosphere and ionosphere, as well as the temperature and bulk velocity of the gas and the homogenous and inhomogeneous reactions of the gas and ions in the dusty cometary atmosphere and ionosphere. The target comet after eight years of cruise was comet 46P/Wirtanen, a small JFC with just about 1 km diameter However, an Ariane V launch failure shortly before the scheduled launch of Rosetta made a launch in January 2003 impossible and 46P could no longer be reached. The new target, 67P/Churyumov-Gerasimenko needed a March 2004 launch and 3 instead of only 2 Earth flybys, making the mission longer. After passing by Mars and the two asteroids Steins and Lutetia, Rosetta finally reached its target in early August 2014 at a heliocentric distance of 3.8 au (see figure 2 for the orbit). After a three-month period of close inspection of the comet down to 10 km from the surface, the lander Philae was released on November 12, 2014. Unfortunately, the harpoons of the lander failed to anchor Philae on the nucleus. This led to extended hopping and a final landing on the winter hemisphere. While not all science goals of Philae could be achieved due to its unfavorable location and attitude, Philae nevertheless worked for 60h sending back invaluable observations.

Rosetta meanwhile continued to orbit the comet at different distances from the nucleus. As the comet became more active closer to the Sun, Rosetta had to draw back to safer distances due to interference of cometary dust with the star trackers. At perihelion the Rosetta –nucleus distance was ~300 km. On the post-perihelion part of the orbit, with heliocentric distances increasing from 1.25 to 4 au, Rosetta slowly went closer to the comet surface. It reached just a few kilometers from the comet in May 2016, and even closer in the last part of the mission, where the minimum distance between the spacecraft and the nucleus surface was <2 km. On Sept. 30, 2016, at almost 4 au from the Sun, Rosetta finally soft-landed on the nucleus, ending the mission as communication was no longer possible.

## 2.1 Comet 67P/Churyumov-Gerasimenko

The target comet of the Rosetta mission, comet 67P/Churyumov-Gerasimenko is a Jupiter family comet with a period of 6.45 y, an aphelion distance of 5.6829 au and a perihelion distance of 1.2432 au. It was detected in 1969 by S. Gerasimenko and K. Churyumov. It acquired its current orbit in 1959 after a close encounter with Jupiter, but is believed to have been in the inner solar system with a perihelion distance < 5 au for several thousand years (Maquet, 2015). However, due to many close encounters with Jupiter its history was chaotic, that means no firm orbital parameters exist before 1923 when it had an earlier encounter with Jupiter.

A picture of 67P is shown in fig. 3. Its largest dimension is 4.1 km. The gravitational point mass is GM = 666.2 ± 0.2 $m^3s^{-2}$, with a mass m = (9982 ± 3) × $10^9$ kg upon arrival at the comet. The average bulk density of the nucleus is 533 ± 6 $kgm^{-3}$ (Pätzold et al., 2016). The nucleus has a porosity of 70-80%. The bi-lobed shape could be a relatively recent feature (Jutzi et al., 2017) whereby the structure can be explained by low-velocity collisions, but there are also authors who claim that the shape and morphology of 67P are compatible with a primordial rubble pile (Davidsson et al., 2016). The bi-lobed shape and the low albedo of 5.9 ± 0.2% at 550 nm (Sierks et al., 2015) are quite typical for the comets which have been visited previously by spacecraft.
The rotation axis is tilted by 52° with a right ascension of 69.3° and a declination of 64.1° (Sierks et al., 2015). Due to non-gravitational forces, its spin period changed. It was 12.4h on arrival of Rosetta and was just above 12h at the end of the mission two years later.
The peculiar shape coupled with the tilted spin axis makes for two very different hemispheres and complicated insolation and dynamics. The northern hemisphere experiences a long (~5.5 y), but cold summer through aphelion, while the southern hemisphere has a short (~10 months), but hot summer through perihelion. The orbit is not symmetric about perihelion, with the inbound equinox in May 2015 at ~1.7 au and the outbound equinox in March 2016 at 2.7 au. The two hemispheres show very different morphology. While the north is dust covered with a rough morphology, the fine dust is mostly lacking in the south. The southern hemisphere can lose up to 10 m per orbit in active areas due to sublimation while the north contributes only about 1/3 to the total mass loss. Self-illumination in concave regions enhances the energy input and hence erosion. The early activity observed from the neck are probably due to this self-illumination and to material which was transported during the previous perihelion passage from the south to the north (Keller et al., 2015).

## 2.2 The highly variable coma between 3.8 au inbound through perihelion and outbound to 3.8 au

**Volatiles:** Not only does the nucleus show morphological heterogeneity, but the coma as well. On arrival at the comet in fall 2014, outside of 3 au from the Sun, the coma was quite heterogeneous, with water coming mostly from the northern, illuminated hemisphere, while the abundance of $CO_2$ was much higher over the southern, at that time winter hemisphere (Hässig et al., 2015). CO followed neither water nor $CO_2$, but was much more evenly distributed over the nucleus surface (Läuter et al, 2018). The ratio of most of the observed species in the coma relative to water was higher over the southern hemisphere (Le Roy et al., 2015). A similar behavior for water and $CO_2$ has been observed for comet Hartley 2 with the EPOXI mission (A'Hearn et al., 2011). Like 67P, Hartley 2 was also bi-lobed. For Hartley 2, $CO_2$ was responsible for the ejection of dust from the short end of the comet, water, which was attributed to water ice grains and fallback material, was coming from the neck.

The outgassing of 67P was monitored constantly by the ROSINA suite, and very regularly by the infrared sensor VIRTIS (Coradini et al., 2007) and the microwave instrument MIRO (Gulkis et al., 2007). While ROSINA measured local density at the location of the spacecraft, the other two instruments derived column densities mostly above the limb. In both cases models are necessary to derive production rates. Figure 4 shows some typical periods for the measured local density for the four main volatile species in the coma, corrected for the comet – Rosetta distance ($1/r^2$).
In the first few months of the mission, the activity of the comet increased only slowly although the heliocentric distance decreased from 3.8 au to 2.2 au by March 2015. The main species is clearly water. The time modulations over a few days are due to the spacecraft position above the nucleus (winter vs. summer). A diurnal variation (~12h) is also observed, which is due to the irregular shape of the comet as the area facing Rosetta changes with the rotation of the comet. A comparison of the production rates derived by the different instruments is found in Hansen et al. (2016). Global production rates derived from different instruments (in situ mass spectrometry, infrared, microwave, and plasma sensors) agree quite well outside of the perihelion, while during perihelion there is more scatter of up to a factor 4. This is due to the fact that Rosetta was close to the comet and no instrument was observing the entire coma. This therefore needs a lot of modelling. Measurements of the water production rate from the LAICA camera on board the micro spacecraft PROCYON using Ly-Alpha (Shinnaka et al., 2017), which encompass the whole coma, are at the upper limit derived from in situ measurements, but still compatible with ROSINA derived production rates. While most of the outgassing is explained by illumination driven sublimation, which means that the whole nucleus is active, some variations in activity over the nucleus are needed to explain observations (Fougere et al., 2016; Kramer et al., 2017).

After the first equinox in May 2015, water and $CO_2$ were released from the same regions in the south of the comet and all species were correlated. By the end of June 2015, the comet started to show frequent short lived (<1h) outbursts (Vincent et al., 2016). The active spots identified from modeling of ROSINA data coincide with the foot points of many of the observed outbursts (Kramer et al., 2017). The driving gas for these outbursts was $CO_2$. After the second equinox in March 2016, the water activity followed the subsolar latitude while $CO_2$ remained active (see the third panel of fig. 4).

From Figure 4 it should be apparent that, depending on the heliocentric distance, pre- or post-perihelion, depending on the spacecraft position over the nucleus, relative abundances of species have considerable variation. This abundance of information makes the question “what is the nucleus bulk abundance” hard to answer. While ROSINA was able to observe the most abundant species like $H_2O$, CO, $CO_2$ over the entire mission, for less abundant species this was not the case. It was found that the period of May-June 2015 is probably the most representative period for the nucleus bulk abundances. The reason is that the comet was inside of 2 au, with all species including water coming predominantly from the southern hemisphere (summer hemisphere). During this time, species correlate quite well independent on the spacecraft position. Closer to perihelion, the frequent outbursts changed relative abundances significantly as they seem to have been driven by $CO_2$ and not water. This is seen in figure 4 by the strong $CO_2$ peak around perihelion. After the peak activity near the end August 2015, water steeply decreased, while $CO_2$ and CO had a much shallower slope. By spring 2016, $CO_2$ was the dominant species in the coma. Selection criteria for choosing the period for bulk abundance measurements are discussed in detail in Calmonte et al. (2016).

# 3 From nucleus composition to coma composition and vice versa

In order to explore cometary chemistry and comet origin, the best way would be to look deep into the nucleus interior, measure composition and its heterogeneity, physical parameters like temperature, tensile strength, porosity, mineralogy of dust and its structure. The lander on Rosetta was equipped with a drill and some measurement capabilities to at least explore the near subsurface material. Due to the unforeseen attitude of Philae at its final landing spot, it was not able to drill and bring up material from the subsurface for analysis. This affected mostly the two mass spectrometers COSAC and Ptolemy. The only material these two instruments could analyze was dust from the first landing site raised by the feet of Philae. Therefore, there are, thus far, no direct measurements of true original nucleus material.

An attempt was made by Deep Impact to look at the nucleus interior by impacting a projectile into comet Temple 1. All parent and daughter fragments observed except ethane were found to be in the same proportions relative to water as in the ambient outgassing, so that fractionation by sublimation, if existing at all, has to be small (A’Hearn, 2011).

All measurements by ROSINA and all measurements for other comets concerning composition of ices were made in the comae. In order to derive nucleus bulk abundances from coma measurements, several different processes have to be taken into account. In the case of 67P, the gas was very tenuous during the entire mission, even during perihelion. The water production rate reached ~$4\times10^{28}$ molecules/s. For larger and much more active comets like 1P/Halley or Hale-Bopp, production rates are at least two orders of magnitude higher. Collisions of molecules and therefore chemistry in the coma in the case of 67P are insignificant most of the time. For other comets, especially in the case of the very active Hale-Bopp and Hyakutake, this is no longer true and chemistry in the coma has to be taken into account. The gas escaping from the nucleus has a velocity between 400-900 m/s (Tzou et al., 2019). As Rosetta was almost always close to the nucleus (<300 km), photodissociation and chemical reaction rates (e.g. Heays et al. 2017, A&A) also do not play a major role. Rates for most of the observed species at these distances from the Sun are much longer than the time needed for the gas to reach Rosetta. Therefore, we expect little difference between nucleus composition and coma composition. This is especially true if we look at isotopic ratios. In contrast to in situ observations, remote sensing observations were very often done over a large radial distance from the nucleus and photodissociation and chemistry has to be taken into account because molecules may have been released long before they are measured.

It is known that some species show a distributed source (e.g. Cottin & Fray, 2008). These species are not directly released from the nucleus, but rather from dust grains. Once in the coma, dust grains can become very hot (Kolokolova et al., 2004). This makes sublimation possible of species that have sublimation temperatures higher than e.g. water. Such distributed sources were detected by the neutral mass spectrometer NMS on Giotto for formaldehyde (Eberhardt, 1999) and subsequently by remote sensing for several species like e.g. $CH_2O$, CN, $NH_3$ (for a summary see dello Russo et al. (2016) and references therein). With Rosetta it was hard to detect distributed sources because the spacecraft did not change its radial distance from the comet over short periods of time. One example where it was possible to observe this effect, was during a fast flyby end of March 2015 with a high amount of dust present. This allowed to follow the glycine abundance, where the density with cometocentric distance r did not follow a $1/r^2$ law (Altwegg et al., 2016). However, ROSINA-COPS (COmetary Pressure Sensor) detected sometimes rapid density increases due to volatile and semi-volatile species sublimating from dust grains (Tzou et al.,2019). Some species could only be observed when the COPS dust counts were high, showing a clear correlation between these species (e.g. $S_3$) with dust.

# 4 Isotopologues

Isotopologues are the best link between comets and presolar material. Isotopic ratios directly reflect the source of the material in the ISM. They are the fingerprints of supernovae, AGB stars, neutron star mergers, etc. While their overall isotopic ratios cannot be changed by chemistry, their specific ratios in molecules can well depend on chemical formation pathways. Isotope fractionation can also occur in cold interstellar clouds, driven by the difference in zero-point vibrational energy of the isotopologues. This difference is largest for light elements like hydrogen / deuterium, where the relative mass difference is highest. Thus, isotopologues of different molecules also tell us about physical and chemical boundary conditions during the formation of the respective molecules.

## 4.1 Water and its isotopologues (HDO, $D_2O$, $H_2{}^{17,18}O$)

### 4.1.1 HDO

One of the most important result of the Giotto mission to comet Halley in 1986 were the measurements of D/H in water by two independent instruments, the Neutral Mass Spectrometer (Eberhardt et al., 1995) and the Ion Mass Spectrometer (Balsiger et al., 1995). Both results yielded D/H ~ $(3.2\pm0.4) \times 10^{-4}$, a value which is twice the terrestrial value (Vienna Standard Mean Ocean Water (VSMOW)). Results from remote sensing for the bright OCC's in the 1990 gave very similar values. This made models for a cometary delivery of the terrestrial bulk water from comets, very unlikely. However, in 2011 Hartogh et al. (2011) reported a terrestrial value for D/H in the first measurements from a JFC, Hartley 2. This revived the idea that JFC's could be the source of terrestrial water. At that time, dynamical models stated that JFC's had been formed further from the Sun, in the vicinity of Neptune's current orbit, while OCC's were formed closer to the Sun. Models predicted higher D/H values with increasing distance from the Sun due to colder temperatures. These models had to be revised following the Hartley 2 result. There exist several different explanations for this inverse behavior of the D/H ratio with distance. One explanation was high temperature gas-phase chemistry in the inner disk (Thi et al. 2010. Another explanation was related to transport of material between the inner and outer Solar System (Walsh et al., 2011). In more recent years, with more cometary D/H values available, it became clear that D/H varies considerably among comets. Measurements of D/H in 67P by ROSINA yielded a very high value of $(5.3\pm0.7) \times 10^{-4}$ (Altwegg et al., 2014). This first value, measured early in the mission outside of 3 au was later confirmed (Altwegg et al., 2017) for post-perihelion periods. This high value rules out fractionation during sublimation because the comet lost several meters of its surface during its orbit, revealing pristine material. Meanwhile, two more D/H values were measured in OCC's, namely in comet C/2012 F6 (Lemmon) and C/2014 Q2 (Lovejoy) (Biver et al., 2016). While comet Lemmon has an even higher D/H value than 67P, comet Lovejoy seems to have an almost terrestrial D/H. The latter result however is disputed by a different measurement (Paganini et al., 2017).

Figure 5 shows an overview of the D/H values in cometary water that are known to date. In Figure 5, it is clear that variations among comets are large. No differentiation between comet families is seen. The lack of variation probably rules out the notion that Oort cloud and Jupiter family comets originated from different places in the protoplanetary disk. Comets probably formed over wide range of radial distances from the young Sun. Their reservoirs (Oort cloud or scattered disk) are the result of their dynamical history, of how they were scattered by the giant planets. Their D/H in water may still represent their place of origin, which would mean that e.g. Hartley 2 formed closer to the Sun than 67P, although they both ended up in the scattered disk.

The average D/H in water of all 10 comets measured so far is $3.6 \times 10^{-4}$. This is more than twice the terrestrial value and is a clear indication that the Earth acquired its bulk water neither from OCC's nor from JFC's. This leaves planetesimals in the early phase of solar system formation as source of the terrestrial water. These bodies have a D/H much closer to our terrestrial standard mean ocean water (SMOW, $1.5\times 10^{-4}$) which, combined with a late veneer of asteroids (Morbidelli et al., 2000), could explain the terrestrial water. Late veneer here defines the last stage of Earth accretion from planetesimals with carbonaceous chondrite-like material which then did not equilibrate with the Earth metallic core and is thought to be about 1% of the total Earth mass (Morbidelli & Wood, 2014). Carbonaceous chondrites (CI and CM) have a mean D/H ratio in their hydroxylated minerals similar to the Earth and are quite water rich of up to 15 wt-% (O'Brien et al., 2014; Robert, 2003). It could also mean that the Earth has preserved enough water in its mantle to replenish dried-out oceans after e.g. the Moon forming impact (e.g. Hirschmann, 2006).

#### 4.1.2 $D_2O$

D/H in cometary water is clearly compatible with values measured in star forming regions and the ISM (see Figure 5) although D/H in these interstellar sources varies from $3\times10^{-4}$ to several % , with some very high values associated to cold gas-phase chemistry. The mechanism for D/H values in water of several times $10^{-4}$, i.e. enhanced compared with the ISM, is based on dust grain chemistry and has been modeled by Cazaux et al. (2011), Taquet et al. (2014) and Furuya et al. (2016; 2017). Furuya et al. show that $H_2O$ is primarily formed in the early, low density stages of molecular clouds whereas HDO and $D_2O$ production is more effective in the later, very cold high density phase (Fig. 1). Furuya et al. followed the HDO/$H_2O$ ratio from the pre-solar stage through the collapse of the cloud and formation of the disk and concluded that this ratio (starting value ~$9 \times 10^{-4}$) was mostly preserved. One result of this modeling was a high $D_2O$/HDO ratio ($4.9\times10^{-3}$) in pre-solar water ice. This high value is within a factor of 2 of the observationally derived values toward the Class 0 protostar NGC 1333-IRAS 2A (Coutens et al., 2014). The high D/H in cometary water alone does not mean that pre-solar water survived the solar system formation without changes. Even though the bulk of the water ice is delivered without alteration, modeling shows that some fraction of ice is affected en route from cloud to disk but that upon sublimation and dissociation and reformation / recondensation, the HDO / $H_2O$ is mostly preserved.

ROSINA was able to measure $D_2O$ during several phases of the mission (Altwegg et al., 2017). The ratio $D_2O$ / HDO = $(1.80 \pm 0.9) \times 10^{-2}$ and therefore f ~ 17. This high value means that most of the cometary water in 67P was probably inherited as ice from the pre-solar stage. Cometary ice of 67P therefore never was inside the water-snow line in the protoplanetary disk. Thus, it is very likely that minor species that were embedded in water ice also never sublimated and thus preserved their pre-solar abundances. Other observations from ROSINA (discussed below) are also consistent with this cometary ice formation beyond the water-snow line.

### 4.1.3 $H_2{}^{18}O$

Oxygen is an abundant element, in volatile species as well as in refractories. It is a light element, so that its three stable isotopes are subject to mass dependent fractionation effects. Also, it undergoes chemical reactions in which non-mass-dependent isotope fractionations occur. Its cosmic abundance causes it to occur simultaneously in two cosmochemical reservoirs: as gas composed mostly of $H_2O$ and CO / $CO_2$ and as solid (oxides and silicates). As part of the refractories, it mostly avoided isotopic homogenization in the interstellar medium and in the early solar system.

Oxygen isotopic composition of our solar system seems then to be the result from mixing two isotopically distinct nebular reservoirs, one $^{16}O$-rich and the other $^{17,18}O$-rich (Sakamoto et al., 2007). The $^{16}O$- rich reservoir is mainly preserved in chondrites which are within a few per mil of the terrestrial SMOW standard. The largest bulk anomalies for the $^{16}O$- rich reservoir are seen in CI chondrites. These anomalies are ~10 ‰/amu, along the terrestrial mass fractionation line (see Hoppe et al. (2018) and references therein). Larger anomalies of up to 60 ‰ enrichment in $^{16}O$ along a line with slope ~1 are seen in CAI's and hibonite grains, which therefore represent the high $^{16}O$, low $^{17,18}O$ – end members of the fractionation line. These $^{16}O$ enrichments are compatible with measurements for the Sun (Mc Keegan et al., 2011).

The low $^{16}O$, high $^{17,18}O$ end-member of this terrestrial fractionation line is COS, cosmic symplectite, a material consisting of aggregates of nanocrystalline iron sulfide and magnetite, identified in the ungrouped carbonaceous chondrite Acfer 094. This seems to represent the $^{17,18}O$-rich reservoir, with large enrichments of about 200 ‰ in $^{17}O$ and $^{18}O$ (Sakamoto et al. 2007). This material is believed to have formed by oxidation of Fe, Ni metal and sulfides by primordial $^{16}O$-poor water in the Solar System. So far, this reservoir represents the best proxy for primordial water (Hoppe et al. 2018). The high $^{17,18}O$ enrichment is explained by self-shielding during the ultraviolet photodissociation of CO (Bally & Langer 1982; van Dishoeck & Black 1988) which leads to an enrichment in $^{17,18}O$ relative to $^{16}O$ atoms, used to form water. According to leading self-shielding models (Sakamoto et al. 2007; Yurimoto & Kuramoto 2004), primordial water is predicted to be enriched in $^{17,18}O$ by 5% to 20% compared to terrestrial water.

Measurements in Stardust samples from comet Wild 2 yielded a $^{16}O/^{18}O$ value very close to the terrestrial fractionation line for large dust grains (>2μm) and quite diverse values for smaller grains (Ogliore et al., 2015) with $^{18}O$ enrichment. They interpreted this result as large grains coming from the inner solar system, exhibiting an equilibrated oxygen isotopic ratio, whereas smaller grains, coming from the outer solar system, preserved their interstellar diversity.

Looking at the entire data set from ROSINA where the signal was high enough to detect $H_2^{18}O$ and $^{18}OH$ and correcting for all known systematic uncertainties, the $H_2^{16}O/H_2^{18}O$ derived is 445 ± 35 (1 $\sigma$ error) (Schroeder et al., 2018). The terrestrial value is 498.7 ± 0.1 measured by Baertschi (1976), while the solar value measured in the solar wind by McKeegan et al. (2011), has a $^{16}O/^{18}O$ ratio of 530 ± 2. The cometary value is more than 1 $\sigma$ outside of the terrestrial value, and more than 2 $\sigma$ outside of the solar value. For $H_2^{17}O/H_2^{16}O$, the relative enrichment in $^{17}O$ is on the order of 17 % compared to the terrestrial value (Schroeder et al., 2019).

The measurements of 67P with an enrichment of ~12 % in $^{18}O$ and ~17 % for $^{17}O$ are fully compatible with primordial $^{16}O$-poor water as inferred from COS. Values from other comets (Table 1) are partly compatible with terrestrial or even solar isotopic ratios, whereas others are compatible with primordial water (see table 1). So far, most oxygen isotopic ratios in comets suffer from relatively high uncertainties and cover a wide range between 300 and 530. Thus, clear correlation between D/H and $^{16}O/^{18}O$ cannot be derived.

ROSINA was also able to measure $C^{18}O^{16}O$ on m/z = 46 (Hässig et al., 2017). Contrary to water, $CO_2$ showed a $^{16}O/^{18}O$ ratio of 494 ± 8, fully compatible with the terrestrial value. This value is predicted by the self-shielding models as $CO_2$ is derived from CO, which is left $^{16}O$ rich. Unfortunately, the value for $C^{18}O$ measured by ROSINA has a high uncertainty due to overlap with NO and in addition is partially a fragment of $CO_2$. However, within the uncertainty, the value is compatible with the terrestrial value as well (Rubin et al., 2017). For the dust, an oxygen isotopic ratio $^{16}O/^{18}O$ of 500 ± 30 was determined (Paquette et al., 2018), compatible with terrestrial as well as with solar values.

All these results are fully compatible with self-shielding in the cloud, leaving cometary water rich in $^{17,18}O$ while not affecting $CO_2$ and refractories and confirms a cloud origin of cometary water.

## 4.2 Deuteration in other molecules

Apart from D/H in water, the only D/H measurement in another molecule before Rosetta is D/H in HCN in Hale-Bopp (Meier et al., 1998) with a value of (2.3 ± 0.4) x $10^{-3}$. This is higher than in water by a factor > 4. It supports the interstellar origin of cometary ices. The authors conclude that the observed value of D/H in HCN implies a kinetic temperature ≥30 ± 10 K in the fragment of interstellar cloud that formed the solar system. With in situ mass spectrometry, this value cannot be confirmed for 67P due to overlap of several species on m/z = 28. However, it was possible to get D/H in $H_2S$ and $NH_3$. The D/H ratio in $H_2S$ is very similar to that in water ((0.6 ± 0.3) x $10^{-3}$) while D/H in $NH_3$ is higher ((1.1 ± 0.2) x $10^{-3}$) (Altwegg et al., 2017; Wampfler et al., 2018). There exists just one value for HDS in a solar type protostar IRAS16293 with D/H = 0.1 (van Dishoeck et al., 1995), but several for hot cores: G10.47+0.03 with D/H = 1.4 x $10^{-3}$, G31.41+0.31 with D/H = 4.9 x $10^{-3}$ and G34.26+0.15 with D/H = 2.0 x $10^{-3}$ (Hatchell &Millar, 1999).

The high D/H in $H_2S$ is consistent with hot cores. A high D/H in $H_2S$ is explained by formation of $H_2S$ in hot core ices at a temperature of 60–80 K (Hatchell & Millar, 1999). D/H in $H_2S$ in comets is therefore compatible with dust grain chemistry processes in the pre-solar cloud.

For $NH_3$ there exist quite a few measurements in protostellar cores in low–mass star formation and quiescent regions in the Galaxy. Generally, D/H values, at least in the gas phase are quite high, between 0.007 and 0.1 (Shah et al., 2011). There might, however, be a gas-phase fractionation effect and the values may therefore not be representative of the ice. Chemical models predict such high values (Roueff et al., 2005). The relatively low value for 67P may point to additional gas phase chemistry in the protoplanetary disk, diluting deuterated $NH_3$. It may also reflect ice chemistry in pre- and protostellar clouds. Depending on which part of the cloud is probed ices can have lower deuterated values than gas (see also Taquet et al. 2014).

## 4.3 Carbon and Nitrogen isotopologues

Carbon and nitrogen are very similar in mass, but not in their chemistry. This may be the reason why carbon isotopes and nitrogen isotopes behave very differently in comets and in meteorites. Carbon isotopic anomalies are generally small, whereas nitrogen anomalies can be very substantial. Observations of volatiles in local and galactic molecular clouds hint at a strong diversity in the $^{15}N$ abundance, extending from a factor 2 depleted to a factor 2 enhanced with respect to the terrestrial ratio (Füri and Marty 2015). Even our Earth has a significantly different $^{14}N/^{15}N$ from the Sun (297 and 440, respectively).

For comets, there exist remote sensing results of the $^{12}C/^{13}C$ ratio in $C_2$, CN, and HCN. The values obtained are within a few percent compatible with the terrestrial standard of 89, however with a few small depletions in $^{13}C$ (Bockelée-Morvan et al. 2015, and references therein).

For comet 67P, ROSINA was able to measure $^{12}C/^{13}C = 84\pm4$ in $CO_2$ with a good precision (Hässig et al., 2017) and in CO (86±9) and $C_2H_5$ (84±12) with limited precision (Rubin et al., 2017). $CO_2$, which is probably a product of a reaction of CO, is slightly enriched in $^{13}C$ (~60 ‰ relative to the terrestrial standard and 140 ‰ relative to the Sun (Hässig et al. 2017)). In the ISM, this ratio varies between 25 to >100, depending on the distance from the galactic center (Wilson 1999), with a local value of 68 ± 15 (Milam et al. 2005).

The main nitrogen bearing molecules in the cometary coma are $NH_3$, HCN and $N_2$. The isotopologues of HCN and $NH_2$, which is most probably a daughter species of $NH_3$, have been measured in several comets. Surprisingly, the $^{14}N/^{15}N$ ratio (~140) is very constant over all measured comets. For $^{14}N/^{15}N$ in HCN, CN, and $NH_2$ there are large enrichments in $^{15}N$ (up to a factor of 3) relative to solar (Bockelée-Morvan et al. 2015, and references therein). This is explained by two different reservoirs, one for atomic nitrogen and the other for molecular nitrogen (Hily-Blant et al., 2017). $N_2$ has so far only been detected in the coma of 67P/CG (Rubin et al. 2015) and as ion $N_2^+$ in a few other comets (e.g. Cochran & McKay, 2018, 2018a). Looking at data from very late in the mission, Wampfler et al. 2018 were able to get a $^{14}N/^{15}N$ value for $N_2$ from m/z = 14 and 15, because at that time most of the other nitrogen bearing molecules were no longer sublimating due to the large heliocentric distance. Quite surprisingly, the value in $N_2$ is compatible to all the other cometary $^{14}N/^{15}N$ values which rules out the theory of two N-reservoirs at least for comets. Thus, the high enrichment of $^{15}N$ is still unexplained.

## 4.4 Sulfur isotopologues of $CS_2$, $SO_2$ and $H_2S$

Due to the higher mass of sulfur bearing molecules, there is little expected isotopic fractionation due to mass dependent or mass independent fractionation. Indeed, meteorites are mostly compatible with the standard VCDT (Vienna-Canyon Diablo Troilite) isotopic ratios for sulfur, at least considering their bulk abundances. However, SiC grains from supernovae exhibit relatively large negative $\delta^{33}S$ and $\delta^{34}S$ compared to the standard value (see fig. 6). In Wild 2, several grains could be analyzed for their sulfur isotopes. Most of them are also within the standard value, for $^{33}S$ as well as for $^{34}S$ (Heck et al., 2012). There is one exception, a grain with a negative anomaly of $\delta^{33}S = -57 \pm 17$ ‰, and $\delta^{34}S = -41 \pm 17$ ‰. This could indeed be a fingerprint of pre-solar grains. The COSIMA instrument (Paquette et al., 2017) reported a value of $^{32}S/^{34}S$ compatible with the VCDT standard. For cometary volatiles, several measurements of $^{34}S$ / $^{32}S$ ratios existed before Rosetta. No values for the ratio $^{33}S$ / $^{32}S$ could be derived for cometary volatiles before Rosetta.

ROSINA obtained values for the $\delta^{33}S$ and $\delta^{34}S$ with lower uncertainties in three species, namely $H_2S$, OCS and $CS_2$ (Calmonte et al., 2017). All values are negative, meaning that $^{32}S$ is enriched relative to the heavier isotopes (see figure 6). The bulk values, the weighted means of the three most abundant species are $\delta^{33}S$ = -51 ± 23 ‰ and $\delta^{34}S$ = -41 ± 17 ‰. The values are not compatible with the standard VCDT, but are similar to what has been observed in SiC-C grains from supernovae (for an overview see Hoppe et al., 2018 and references therein) and similar to observations of the ISM (Mauersberger et al. 2004). Within 1 $\sigma$ uncertainties, previously measured values in comets Halley, Hale-Bopp, Lovejoy and Lemmon are compatible with 67P, but also with the standard VCDT due to their much larger uncertainties.

Non-solar isotopic ratios for volatile molecules and values compatible to solar values for the refractories in 67P point to a non-homogeneized protoplanetary disk from which comets formed. This conclusion is fully compatible with the $^{16}O$ poor water of 67P.

## 4.5 Halogene isotopologues (HCl, BrCl)

For the halogens, generally isotopic ratios for Cl and Br are quite constant in chondrites, Vesta, the Earth and the Moon, the Sun and for the protostellar core value in present-day molecular clouds in the local galactic environment (Kama et al., 2015). This constancy seems also to hold for 67P, where the isotopologues of HCl and HBr were detected (Dhooghe et al., 2017). This is not surprising because halogens do not undergo much chemistry and their masses are quite high, such that mass fractionation may not play a big role.

## 4.6 Silicon isotopes

Silicon isotope anomalies are generally small (<1 ‰) for chondrites (Poitrasson 2017) except pre-solar grains, especially SiC grains. 90% of these grains are enriched in heavy isotopes and are believed to be the result of AGB stars of about solar metallicity (Lugaro et al. 2003). SiC grains from SNe, the X and C grains, which make up 1-2 % of all pre-solar SiC grains show large depletion in heavy isotopes (Zinner 2014). For a thorough discussion on Si isotopes in meteorites see Hoppe et al. (2018). There are only a few measurements of Si isotopes in the ISM and they match solar values, but with large error bars (e.g., Tercero et al. 2011; Wilson and Rood 1994).

Isotopic composition of dust in comets has been measured in samples from the Stardust mission to comet Wild 2, as well as in situ during the Giotto mission to comet Halley in 1986 and by COSIMA for 67P. However, none of these measurements targeted Si isotopes for various reasons. Early in the Rosetta mission, when the activity of the comet was still quite low, the solar wind impacted the surface of the comet, sputtering dust. ROSINA measured some refractory components like K, Mg and Si (Wurz et al., 2015). ROSINA determined three Si isotopes: $^{28}Si$, $^{29}Si$ and $^{30}Si$. The isotopic ratios were $^{28}Si$ / $^{29}Si$ = 23.0 ± 2.6 and $^{28}Si$ / $^{30}Si$ = 38.0 ± 5.6. These values are higher than the solar values (19.68 and 30.77, respectively) and can only be explained by a strong supernova contribution to the protoplanetary disk (Hoppe et al. 2018).

## 4.7 Noble gas isotopes

The Rosetta mission provided the first ever possibility to measure the abundance of the three noble gases Ar, Kr, and Xe in a cometary coma, as well as their isotopic ratios. As noble gases undergo very little to no chemistry, the isotopic ratios therefore have to reflect their nucleosynthetic origin, modulated at most by some mass fractionation. All three noble gases studied have relatively high masses, therefore the mass fractionation effects are estimated to be small.

For Ar the measured isotopic ratio is $^{38}Ar$ / $^{36}Ar$ = 5.4 ± 1.4 (Earth: 5.3; solar wind: 5.5) (Balsiger et al., 2015). Unfortunately the precision of this measurement, due to overlapping species at mass 38 and 36, is not sufficient to rule out any deviation from terrestrial, from solar argon or from any of the specific components found in meteorites. No $^{40}Ar$ was detected, which is not surprising as this isotope is not primordial.

Kr has five stable isotopes, $^{80}Kr$, $^{82}Kr$, $^{83}Kr$, $^{84}Kr$ and $^{86}Kr$. All of them were clearly seen in the mass spectra of ROSINA in May 2016, when the spacecraft was ~7 km from the nucleus center (Rubin et al. 2018). $^{80}Kr$, however, has a large error bar due to overlap with $C^{34}S_2$. All Kr isotopes, relative to the most abundant $^{84}Kr$, are solar within the error limits, with the exception of a slight depletion of $^{83}Kr$ (see table 2).

Xenon has nine stable isotopes, $^{124}Xe$, $^{126}Xe$, $^{128}Xe$, $^{129}Xe$, $^{130}Xe$, $^{131}Xe$, $^{132}Xe$, $^{134}Xe$ and $^{136}Xe$. All of them except the two lightest ones could be observed in the same measurement sequence with Kr and its isotopes (Marty et al., 2017). Normalized to $^{132}Xe$, there is an enrichment in $^{129}Xe$, which can be attributed to the radioactive decay of $^{129}I$. Therefore, more $^{129}I$ was available when icy grains formed than at the start of the solar system formation when meteorites formed. The higher mass $^{134}Xe$ and $^{136}X$ are clearly depleted relative to $^{132}Xe$ by 35 % and 60 %, respectively. This depletion cannot be explained by mass dependent fractionation and has to be a nucleosynthetic effect.

Meteorites contain trapped noble gases and those produced in situ by spallation or radioactive decay. For a review see Ott (2014). The trapped planetary noble gases are classified into different components according to their distinct isotopic patterns (Q, P3, HL, P6, G, N). Except Q, all components are believed to have a presolar origin and are the products of a mixture of different progenitor stars with their specific history of nucleosynthesis. The Xe-isotopic pattern of 67P/CG differs from the patterns of the trapped planetary noble gas components. It is neither similar to the Q component nor to the components associated with pre-solar grains. However, the Xe pattern corresponds quite well to a mixture of s-process Xe and two r-process end-member compositions identified by Gilmour and Turner (2007). The Kr isotopic ratios were near solar with the exception of $^{83}Kr$ (slightly depleted).

To explain the quasi-solar Kr isotopic ratios and the clearly non-solar Xe isotopic ratios, a mixing scenario of a “normal” component N-type Kr and N-type Xe with an exotic, s-process rich component called G-type Kr and G-type Xe, respectively can be used (Rubin et al., 2018). However, this scenario would mean that the G-component is much more prominent in Xe than in Kr. Studies of pre-solar material have shown that indeed Xe is commonly enriched relative to Kr in s-process end-members. This enrichment is attributed to chemical fractionation like differences in the behaviors of Xe and Kr during expulsion from stellar envelopes, preferential ionization of Xe relative to Kr, and/or selective implantation of Xe into dust.

What is interesting is that by mixing (22 ±5) % of cometary Xe with chondritic Q-Xe it is possible to recreate the primordial atmospheric U-Xe component on Earth, postulated 40 years ago by Pepin (Pepin & Phinney, 1978; Pepin, 2000). This allows to estimate the amount of water delivered to the Earth by comets which is on the order of < 1% (Marty et al., 2017). This also gives us a handle on the importance of cometary delivery of organics to the early Earth (see below).

## 4.8 What can we learn from isotopes

Table 2 gives a summary of what we know about isotopologues in cometary ice based on the subchapters in this chapter. Measured values in 67P and, if available, average values in other comets are compared to solar values or, if not available, standard values from meteorites.

Isotopes on one hand give a fingerprint of the precursor stars that fed the protoplanetary disk. This is especially true for species like noble gases that do not undergo chemical reactions. On the other hand species that undergo chemical reactions in cold interstellar clouds are subject to chemical fractionation. Chemical fractionation depends on the physical parameters like temperature, density, time available, UV radiation and cosmic rays.

Looking at the isotopic ratios of the noble gases, it is clear that 67P did not acquire the same complement of isotopes as the Sun. The enrichment of $^{129}Xe$ may be due to an excess in $^{129}I$ at the location the cometary ice formed. The depletion in $^{134,136}Xe$ points to a different mixture of nearby remnants of stars at the location of comet formation than the average of the solar nebula. Comet impacts during the late veneer / late heavy bombardment may then have brought this xenon component to the early Earth. It most probably means that the protosolar nebula was not well mixed. The Si isotopic ratios as well as the sulfur isotopic ratios support this conjecture.

From deuterated species, especially the abundant $D_2O$, we conclude that water ice formed in the pre-solar cloud and survived the collapse of the solar nebula as ice. We also conclude that the notion that different comet families formed at different distances from the Sun is probably not correct. The large differences in D/H in cometary water, independent of the comet family points more to a scenario where comets formed over a relatively large region in the protoplanetary nebula and were later scattered either in an Oort cloud orbit or ended up in the scattered disk.

Looking at 67P and comparing it with other comets, we find little differences, except in the already mentioned D/H. Most isotopic ratios measured in comets are compatible within uncertainties with the measurements in 67P, although uncertainties from remote sensing are generally higher.

The mystery of a constant $^{14}N/^{15}N$ in comets over all molecules including $N_2$ remains yet to be solved.

Other isotopic ratios like sulfur seem to be correlated with ice chemistry, again pointing to a pre-solar origin of the molecules. Oxygen isotopes in water and $CO_2$ are well in line with self-shielding models in the pre-solar nebula, again pointing to pre-solar chemistry. Overall, isotopic ratios in 67P are in line with what has been measured in star forming regions and hot cores, strengthening the link between comets and the ISM.

# 5 Molecular composition

In this section we look at molecular abundances of species found in cometary comae, especially in 67P. As shown in figure 1, the material that finally makes up comets has passed through several stages, each with its own physical characteristics. These characteristics determine what species are formed or changed at which stage. Molecular clouds prior to their collapse have low densities and cold temperatures. Star forming regions have higher densities and higher temperatures close to their protostar. Protoplanetary disks have diverse conditions. In the mid-plane, there are low cosmic rays and UV, whereas out of the midplane, UV dissociation and ionization sets in. The amount of radial and vertical mixing in the protoplanetary disk is still under debate. An overview of the processes in star forming regions through the collapse and the formation of planetesimals is given in e.g. van Dishoeck & Blake (1998) and Johansen et al. (2007). By comparing what is found in comets and what is found in star forming regions, some of the processes in the protoplanetary disks are better constrained. A list of detected molecules in comets is found in table 4.

## 5.1 Highly volatile species: CO, $CO_2$, $N_2$, $CH_4$, and $O_2$

Looking at highly volatile species indicates the formation temperatures of comets as well as their evolution in the last 4.6 billion years. CO and $CO_2$ are regularly observed in comets with remote sensing. Fig. 4 clearly shows that in 67P the ratio between water, $CO_2$ and CO is highly variable. While outside of 3 au inbound, water seemed to be anti-correlated with $CO_2$, being high over the northern (summer) hemisphere, CO showed little variation with latitude. Around the highest activity at the end of August 2015, all three species had higher densities over the southern hemisphere, at that time in summer, but still, $CO_2$ showed larger variations than $H_2O$ or CO. Outbound, outside of 3 au, $H_2O$ followed the subsolar latitude, which was at that time around 5° north, while CO and $CO_2$ were very well correlated and still came from the south. When comparing comets, these intrinsic variations of relative abundances, which are related to the shape and rotation axis of the comet, to fallback material (Keller et al., 2015), and the heliocentric distance have to be taken into account. The probably best values for the nucleus bulk relative abundances are from May / June 2015, with the southern hemisphere being the summer hemisphere, well inside of 2 au and before the onset of the summer firework (Vincent et al., 2017), describing frequent short lived outbursts of the comet (see also chapter 3.8)

67P, with a relative abundance of $CO/H_2O$ ~ 1.7 %, compares well with the average abundance of 1.6 ± 0.7 % for other JFC's (dello Russo et al., 2016). 67P does not compare well with the average abundance of 6.1 ± 1.6 % for OCC's (dello Russo et al., 2016). $CO_2$ is difficult to measure with remote sensing. Therefore 67P is the first JFC where it has been measured. With an abundance of ~4.5 % it is rather typical for comets, which have a $CO_2/H_2O$ ratio between 2.5 and 12% (Mumma & Charnley, 2011).

$CH_4/H_2O$ is 0.34 % (Schuhmann et al., 2018), which again is at the low end of the range seen in other comets, but compatible with measurements in two JFC's, P/Encke (0.34%) and Temple 9 (0.54%). The average abundance for Oort cloud comets is almost a factor 3 higher than for JFC's (dello Russo, 2016 and references therein).

$N_2$ was detected for the first time in a comet by ROSINA (Rubin et al., 2015). The abundance for $N_2$ is $(8.9 \pm 2.4) \times 10^{-4}$ relative to water. This abundance is not enough to solve the problem of the N-deficiency in comets already noted by Geiss et al. (1988) for comet Halley. The explanation for the deficiency in N at that time was that formation temperatures of comets were too high to freeze out $N_2$ or that comets have lost their $N_2$ along the way from the Oort cloud to the inner solar system. With the detected $N_2$ we can at least conclude that some $N_2$ was implemented in comets and not all of it was lost, which points to a formation temperature of the comet <50 K (Rubin et al., 2015).

$O_2$: Probably the most surprising result of the ROSINA instrument at 67P was the detection of abundant $O_2$ (Bieler et al., 2015). $O_2$ is the third or fourth most abundant species in the coma (see fig 4), very close to CO. Although $O_2$ is very volatile, it follows much more $H_2O$ and not CO or $CO_2$ over the entire mission with an abundance of (1.8 ± 0.4) % relative to water. This value is somewhat lower than the value derived by Bieler, reflecting a larger dataset used in Läuter et al., 2018. This detection triggered a re-analysis of the data from the Neutral Mass Spectrometer NMS on Giotto. Indeed, Rubin et al. (2015b) found strong indications that also comet 1P/Halley showed $O_2$ in its coma, with a very similar abundance relative to water as 67P. Several explanations on the formation of $O_2$ have been published since its detection. Interstellar $O_2$ gas is difficult to detect as the transition lines are weak, but nevertheless low observational upper limits exist. So far, a very deep search of $O_2$ has been made in only one potential precursor of a solar-like system protostar, NGC 1333–IRAS 4A ($O_2/H_2$ $< 6 \times 10^{-9}$; Yildiz et al. 2013). The most plausible formation scenario is by Taquet et al. (2016; 2017), who postulate gas-grain chemistry in the dark cloud. This scenario requires a slightly warmer cloud than normal (~20K) in order to also avoid overproducing $HO_2$ and $H_2O_2$. These authors rule out chemistry during formation of the protosolar disk or due to luminosity outbursts in the disk. Disk models cannot explain the strong correlation between $O_2$ and $H_2O$ in comet 67P/C-G together with the weak correlation between other volatiles and $H_2O$. However, primordial $O_2$ ice can survive transport into the comet-forming regions of the protoplanetary nebula, if water ice never sublimates, which is in agreement with the isotopologues of water.

Another explanation is given by Mousis et al., 2016. This explanation is based on radiolysis of water ice in the pre-solar ice grains. Radiolysis means dissociation of molecules by ionizing radiation, e.g. electrons, cosmic rays, etc.. This process cannot happen in the protoplanetary disc as cosmic rays are not abundant enough. Both mechanisms, but especially radiolysis, produce abundant $O_3$ and $H_2O_2$, which has not been detected by ROSINA. However, $O_3$ might not survive in ice long enough to be detected.

Dulieu et al. (2017) suggested formation of $O_2$ during water ice desorption through dismutation of $H_2O_2$. This process would produce abundant $H_2O_2$, about 4 orders of magnitude higher than observed by ROSINA. Another proposed mechanism is an Eley-Rideal process (Yao et al., 2017), whereby pick-up water ions hit the cometary surface and produce $O_2$. This process is not consistent with observations from the plasma instruments, as $O_2$ is anti-correlated with $H_2O^+$ (Heritier et al., 2018).

Recent attempts to quantify $O_2$ in solar-type protostar IRAS 16293—2422 (Taquet et al., 2018) yielded an upper limit, lower than the value measured in 67P, which may be due to the colder temperatures of the cloud out of which this protostar formed compared with that of our protosun. It appears that the presence of abundant $O_2$ in 67P can only be explained by pre-solar processes, either gas-grain chemistry in the dark cloud or radiolysis in pre-solar ice grains. In the second case the lack of abundant $H_2O_2$ remains to be explained. In order to maintain the correlation with water ice, water ice cannot have sublimated before accretion in the comet.

67P seems generally to be depleted in the highly volatile species ($CH_4$, CO, $N_2$) compared to Oort cloud comets. The low values for $CH_4$ and CO in 67P are in line with other Jupiter family comets (Dello Russo et al., 2016). If we follow the argumentation in chapter 4.1.1. that the D/H ratios in water are variable but do not fundamentally differ between the two comet families, the difference in the abundance of highly volatiles may therefore be a consequence of their dynamical history, rather than of their place of origin. Comets originating in the scattered disk to become Jupiter family comets normally reach the inner solar system by first becoming Centaurs at heliocentric distances between 5 and 50 au. This stage can easily last $10^6$ y. Thermal models show a temperature increase of up to 60 K in the outer few 100 m of Centaurs, leading to potential diffusion and release of highly volatiles (Guilbert-Lepoutre, 2016). On the other hand, 67P is eroding several meters per orbit once in the inner solar system and it is therefore not clear if the depleted layer has already been eroded in the last few 1000 y.

## 5.2 Abundances of noble gases

Before Rosetta, no clear detection of noble gases in a comet was possible. ROSINA was able to detect Ar, Kr and Xe including most of their isotopes (Marti et al., 2017; Rubin et al., 2018) and an upper limit for Ne (Rubin et al., 2018). The noble gases correlate well with $N_2$, somewhat less with $CO_2$ and not very well with water. They could only be observed very close to the comet, which made their detection during perihelion not feasible. Assuming that the correlation with $N_2$ also holds for the perihelion passage, where the coma composition was probably closest to the bulk nucleus composition, Rubin et al. (2018) obtained the following abundances: $Ar/H_2O = (5.8 \pm 2.2) \times 10^{-6}$; $Kr/H_2O = (4.9 \pm 2.2) \times 10^{-7}$; $Xe/H_2O = (2.4 \pm 1.1) \times 10^{-7}$; $Ne/H_2O < 5 \times 10^{-8}$ using $N_2/H_2O = (8.9 \pm 2.4) \times 10^{-4}$ and $^{36}Ar/N_2 = (5.5 \pm 1.5) \times 10^{-3}$.

Assuming water ice to be amorphous these abundances would point to relatively high trapping temperature (≥70 K) as deduced from laboratory experiments (BarNun et al., 2012). This then, however, contradicts e.g. the $CO/N_2$ ratio which requires temperatures ≤50 K (Rubin et al., 2015).

These authors have also speculated that water ice is forming clathrates. Models for clathrates yield higher Xe/Ar abundances than observed. For clathrates or crystalline ice , no applicable lab experiments exist.

There are several facts that affect the relative abundances: lab measurements are done usually for pure amorphous water ice only. The application of these lab results to real cometary ice has been questioned by e.g. Kouchi &Yamamoto (1995) and Greenberg et al. (2017). Addition of $CO_2$ can change the trapping efficiency quite radically, but there are so far no measurements on how this affects the different noble gases. The fact that noble gases correlate much better with $CO_2$ than with $H_2O$ indeed supports the need for more laboratory work with mixed ices. The volatility of Ar is very high compared to Xe, which is quite sticky. This would also favor large ratios of Kr/Ar and Xe/Ar as the upper layers of the comet probably experienced temperatures up to 60K during its transfer from the Kuiper belt to becoming a JFC. This is in line with the depletion for JFC's for $CH_4$ and CO relative to OCC's.

## 5.3 Sulfur bearing species

Sulfur is the 10$^{th}$ most abundant element in the Universe. Its cosmic abundance relative to oxygen is 2.7%. Organic sulfur chemistry resembles oxygen chemistry. Many molecules are very similar, just exchanging oxygen with sulfur like e.g. formaldehyde ($H_2CO$) and thioformaldehyde ($H_2CS$). However, sulfur can form many allotropes, second only to carbon, able to exist from polar $H_2S$ to stable, refractory ring molecule $S_8$.

Measurements of sulfur in the diffuse interstellar medium actually confirm the cosmic value (Savage & Sembach 1996; García-Rojas et al. 2006; Howk, Sembach & Savage 2006; Jenkins 2009) which also means that most of the sulfur is in gaseous form. On the other hand, in dense clouds and star-forming regions, sulfur seems to be highly depleted by about a factor of 1000 relative to its cosmic abundance (Penzias et al. 1971; Tieftrunk et al. 1994).

While dense clouds lack sulfur, our solar system, especially the planets and their moons, show quite abundant volatile sulfur species. $SO_2$ (Toon, Pollack & Whitten 1982) and $S_n$ (n=1-8) (Esposito, Winick & Stewart 1979) are responsible for the reddish color to Venus' fumes. The Jupiter moon Io also shed sulfur from its volcanoes in the form mostly of $SO_2$ (Pearl et al. 1979) as on Earth. Some of the reddish color on Jupiter may originate from $S_8$. For sulfur/hydrogen, the enrichment in Jupiter is up to a factor 2.5 compared to solar values (Atreya et al. 2003).

This puzzle is explained by grain surface chemistry starting with $H_2S$ formed by hydrogenation (e.g. Woods et al., 2015 and references therein). However, $H_2S$ is not detected in the ice in clouds, with an upper limit of $H_2S/H_2O$ of as low as 0.2 % (Jiménez-Escobar & Muñoz Caro, 2011). This led to the hypothesis that $H_2S$ is destroyed in water-rich grain mantles by UV photons, X-rays or cosmic rays. Ice processing experiments have shown that $H_2S$ can be transformed into OCS and $S_2$. By further photolysis or radiolysis S-polymers up to $S_8$ can be obtained as well as organo-sulfurs (e.g. Ferrante et al., 2008; Jiménez-Escobar & Muñoz Caro, 2011). OCS could also be formed by low-temperature surface reactions of CO with S without the need for photolysis or radiolysis.

Many sulfur species have been detected in comets. $H_2S$, SO and CS are regularly observed with remote sensing, with $H_2S$ being by far the most abundant (fig. 7). In 67P, the second most abundant species is surprisingly atomic sulfur (Calmonte et al., 2016). This cannot be attributed to fragments of observed molecules. An excess of atomic sulfur has also been found from UV observations of several comets (Meier et al., 1997). CS is much easier to detect by remote sensing than $CS_2$; however, in situ CS at 67P is clearly a fragment of $CS_2$ and no direct release of CS is observed. $S_2$ in 67P was detected outside of 3 au where the coma was very tenuous. This detection early in the mission rules out formation of $S_2$ by coma chemistry as suggested by A'Hearn et al. (1983) or Saxena et al. (2003). In addition, Rodgers & Charnley (2006) showed that coma chemistry is unlikely to form the relatively high amount of $S_2$ seen in cometary coma. In 67P, closer to perihelion, the relative abundance between $H_2S$ and $S_2$ changes significantly. It increased from the stable 0.1% over much of the time of the mission to several % with large variations during the perihelion passage. This increase suggests that $S_2$ has two sources: a direct release from the ice as $S_2$ and a refractory component related to the polymer $S_n$. Coincidentally, perihelion was also the time when dust impacts on the spacecraft occurred and $S_3$ and $S_4$ were clearly detected by ROSINA (Calmonte et al., 2016). The abundance ratio at this time was $S_3/S_2 \approx 3$ and $S_4/S_2 \approx 1$, although with rather large uncertainties. The detection of $S_3$ and $S_4$ confirms the photolysis / radiolysis pathway proposed previously (Woods et al., 2015).

$S_2$ has a very short lifetime in the gas phase and is destroyed by UV at 283 nm within 250 s at 1 au (De Almeida & Singh; 1986). Heays et al. (2017) give an even shorter lifetime of 130s. Even in the dark midplane, $S_2$ gas would not survive long. This implies that the ice arrived in solid form directly from the interstellar medium in accordance with the doubly deuterated water and the abundant $O_2$. Furthermore, that the ice did not sublimate and re-condense in the protoplanetary nebula.

$H_2CS$ has been detected in the low-mass protostar IRAS 16293−2422 (Schöier et al. 2002). Majumdar et al. (2016) and Drozdovskaya et al. (2018) detected $CH_3SH$ for the same source. This molecule has also been detected in massive star-forming regions such as SgrB2 close to the galactic center (Linke, Frerking & Thaddeus 1979; Müller et al. 2016), in the massive hot core G327.3−6 (Gibb et al. 2000a) and in Orion KL (Kolesniková et al. 2014). The abundance relative to $H_2S$ can be estimated by taking the abundance of $H_2S$ (compared to $H_2$) of $9 \times 10^{-8}$ reported by Schöier et al. (2002) which gives a ratio $CH_3SH/H_2S = 4.4 \times 10^{-2}$ for IRAS 16293−2422, comparable to that of 67P which is $(1.4 \pm 0.4) \times 10^{-2}$.

$C_2H_6S$, together with $CH_3SH$, was detected in Orion KL by Kolesniková et al. (2014). Their estimated $CH_3SH/C_2H_6S$ ratio is ~5 which is similar to 11 ± 4 found for 67P (Calmonte et al., 2016). The two ratios are surprisingly close, even though Orion KL is a high-mass star-forming region. This similarity is in accordance with a common process at work, namely grain surface chemistry in the pre-stellar clouds.

Comparison of 67P sulfur with ALMA Band 7 data for IRAS 16293–2422 B showed that OCS in 67P relative to $H_2S$ is depleted, whereas the ratios of $SO/SO_2$ and $CH_3SH/H_2CS$ are very similar (Drozdovskaya et al., 2018). The differences in sulfur bearing species could be due to the levels of UV radiation during the initial collapse of the systems. These levels may have varied and potentially may have been higher for IRAS 16293-2422 B due to its binary nature, enhancing radiolysis of $H_2S$.

If we look at the overall sulfur abundance in comets and use the partitioning of sulfur between dust and gas derived in comet 1P/Halley, sulfur in comets relative to oxygen is compatible with the cosmic abundance of S/O = 2.7%(Calmonte et al., 2016). The depletion seen in pre-stellar clouds may therefore just be due to the bias of remote sensing as sulfur on dust grains or in the ice is not accessible. The James Webb Telescope may help to shed more light on the missing sulfur in clouds.

## 5.4 Halides and phosphorous

Halides have been detected in the ISM in the form of HCl (Blake et al., 1985) and HF (Neufeld et al., 1997). Ligterink & Kama (2018) searched for HBr and HBr+ in the Orion KL Hot Core. They deduced an upper limit on $HBr/H_2O$ which is a factor 10 below the ratio measured in comet 67P. However, using a chemical network, they conclude that this result is compatible with cometary bromine as most HBr is not in the gas-phase. Cometary HBr is predominantly formed in icy grain mantles not accessible to remote sensing. A search for HCl and HF in comets 103P/Hartley 2 and C/2009 P1 (Garradd) by Bockelée-Morvan et al. (2012) with Herschel resulted in no detection for HCl and marginal detection for HF in comet C/2009 P1. For 67P HF, HCl and HBr and the isotopologues of HCl and HBr were all measured by ROSINA (Dhooghe et al., 2017). Their ratios relative to water are not constant over the mission. De Keyser et al. (2017) explain this by release from halogen-containing water ice mantles on refractory grains. At least for HCl and most probably also for HBr there is strong indication that they could be stored as ammonium chloride ($NH_4Cl$ and $NH_4Br$, respectively) on cometary grains (Altwegg et al., 2019). Table 3 shows the relative abundances for the halides in the coma of 67P, in the Sun and the Earth. While Cl and F are comparable to the solar values from Lodders (2010), the Earth is lacking halogens. All of the halogens in 67P are in line with what is known from chondrites. This similarity probably also confirms that halides correlate with dust grains.

Recently, a study with ALMA of the gas surrounding the low-mass protostar IRAS 16293-2422 has yielded the detection of the first organo-chloride, $CH_3Cl$ (Fayolle et al., 2017). This molecule was thought to be a biomarker as it is abundantly produced on Earth by biological activity. $CH_3Cl$ was revealed in 67P, although only during very dusty periods when HCl was at its highest abundance. This detection yielded a relative abundance $CH_3Cl/HCl = (4 \pm 2) \times 10^{-3}$. Comparing the ALMA detection and the value in 67P the ratio $CH_3Cl/CH_3OH$ for the star forming region of ~ $7 \times 10^{-5}$ and for 67P of (0.007 to 6) x $10^{-4}$ are very similar (Fayolle et al., 2017). This result is a clear indication that $CH_3Cl$ is not necessarily related to biological activities.

Phosphorous is both scarce and ubiquitous in the universe. It is an essential ingredient for life as we know it. Its cosmic abundance is much smaller than that of nitrogen ($[P]/[N] = 4 \times 10^{-3}$), which is in the same valence group. This is due to the complex nucleosynthesis of the P nuclide in the cores of massive stars, leading to a wide distribution of this element through the ISM. A general tendency towards more oxidized phosphorus compounds with chemical evolution proceeds from ISM to clouds to protoplanetary condensed matter phases. Several phosphorous bearing molecules have been detected in the ISM and in clouds: e.g. PN in clouds (Ziurys, 1987), PO in the envelope of the oxygen-rich supergiant star VY Canis Majoris (Tenenbaum et al., 2007), CP (Guélin et al. 1990) and HCP (Agúndez et al., 2007) in the AGB star envelope IRC +10 216. In the solar system, phosphorous is very often in the form of phosphate ($PO_4$), except in the atmosphere of Jupiter where phosphine $PH_3$ has been detected. In meteorites, P is the 13$^{th}$ most abundant element, although it ranks 18$^{th}$ in cosmic abundance. Thus, it is enriched in meteorites as well as in the Earth mantle. In comets, phosphorous is quite elusive. Before the Rosetta mission, there were no detections, neither in the gas nor in the dust. A clear signature on the exact mass of P was detected in the coma of 67P (Altwegg et al., 2016). At that time it was not clear which molecule was responsible for the P in the spectrum. A careful analysis recently revealed that the most probable parent is PO (Rivilla et al., in preparation). The peak in the mass spectra at $m/z = 46.969$ is not unique as it could belong to $C^{35}Cl$ as well as PO. However, this peak correlates well with the peak at $m/z = 30.974$ (P) over time, but not with $^{35}Cl$. Also there is no peak at the corresponding mass for $C^{37}Cl$. For $PH_3$, there is only an upper limit of $PH_3 / PO < 30\%$ due to overlapping masses of $H_2S$, $^{16}O^{18}O$ and $H_2O_2$. For CP/PO the upper limit is at < 3% and for PN/PO at < 10%. No higher oxidation states were found.

These findings are compatible with gas phase chemistry of phosphorous in a cold environment. $P^+$ does not react in the gas phase with $H_2$ nor does $PH^+$ (Thorne et al., 1984). $P^+$ can react with $O_2$, $CO_2$ and $H_2O$ to form the very stable $PO^+$ and with $NH_3$ to form PN, which makes PO and PN the most plausible phosphorous bearing molecules in dark clouds and pre-solar nebula (Millar, 1991) for gas phase chemistry. However, gas-grain chemistry would probably freeze out and hydrogenate phosphorous as $PH_3$. A model by Charnley and Millar (1994) shows that $PH_3$, released in hot cores from grains has a lifetime of $<10^5$ y before being converted mostly into PO and PN. This may explain why no $PH_3$ is detected in hot cores / star forming regions. The detection of PO and the non-detection of $PH_3$ in 67P may therefore be the result of gas phase chemistry or the result of gas-grain chemistry and subsequent destruction of $PH_3$.

## 5.5 Inorganic nitrogen bearing species

Contrary to phosphorous, nitrogen readily hydrogenates to form $NH_3$. Ammonia is the most abundant nitrogen-bearing molecule in comets, with a relative abundance compared to water of 0.8 %. It is readily observed through the daughter products NH and $NH_2$. An overview of all measured cometary values is given in dello Russo et al. (2016). What is interesting about ammonia is that it seems to have a higher abundance for comets coming close to the Sun, showing a distributed source, with ammonia being released from grains in the coma. For 67P, $NH_3$ followed $H_2O$ through the entire mission, similar to $O_2$ (Gasc et al., 2017), although $NH_3$ is much more volatile. It seems to be very well embedded in water ice or be present in a less volatile form than pure ammonia, e.g. as ammonium salt. During dust impacts, ammonia could become two orders of magnitudes higher than the "normal" value, also pointing to a less volatile form of $NH_3$ correlated with dust grains. Work on ammonia is ongoing (Altwegg et al., 2019).

The other inorganic nitrogen-bearing molecule is NO, which was also found in 67P. The release of this molecule is correlated more with $CO_2$ than with $H_2O$. No $NO_x$ with x>1 was found. This is similar to the case of PO. The relative abundance $NO/H_2O$ is ~$2x10^{-4}$ (Wampfler et al. 2018).

## 5.6 Saturated organics

One of the big surprises from the Giotto mission more than 30 years ago was the amount of heavy species detected in the cometary coma (Korth et al., 1986). These species were clearly attributed to organics. In the ensuing 30 years, many complex organic molecules have been found by remote sensing in comets (e.g., ethanol, ethylene glycol, formamide and glycolaldehyde in comet C/2014 Q2 (Lovejoy) and C/2012 F6 (Lemmon), respectively (Biver et al., 2014; 2015). Due to the high sensitivity of the ROSINA instrument and the long period of observation, this list of detected organic molecules has now almost doubled.

### 5.6.1 Aliphatic and aromatic hydrocarbons

Comets contain quite a lot of pure hydrocarbons like methane, ethane, ethylene, etc. They are, however, not easy to detect by remote sensing, because they have zero dipole moments. With mass spectrometry, they are relatively easy to identify, as their mass compared to the integer mass is large due to the many hydrogens attached. Fig. 8 shows the hydrocarbons in 67P (Schuhmann et al., 2018; LeRoy et al., 2015) at ~1.4 au. What is striking is that the abundances of these hydrocarbons up to propane are very similar during the preperihelion part in May-June 2015, which we consider to be the best proxy for the nucleus bulk abundance (see 2.2). However, the relative abundances, especially between $C_2H_6$ and $CH_4$ vary by several orders of magnitude over the mission, $C_2H_6$ being much more abundant than $CH_4$ at large heliocentric distances. This is difficult to explain, but is probably related to how these species are embedded in the dust-ice matrix in the comet. Clearly more work is needed on the desorption processes. Butane and pentane are only seen in conjunction with dust. This may be due to their higher sublimation temperatures. ROSINA clearly detected benzene and the more abundant toluene, a molecule that tentatively has also been detected by Ptolemy on the lander (Altwegg et al., 2017). During dust events, there are also tentative detections of octane, xylene and naphthalene in the mass spectra of ROSINA. This, however, is still work in progress.

### 5.6.2 Other organics

An overview of identified organics by ROSINA is given in table 4b (Altwegg et al., 2017; Schuhmann et al. in preparation). With mass spectrometry, it is not always possible to separate isomers, as they have the exact same mass and differ only in their fragmentation pattern upon electron impact ionization. Species that could be an isomer and species where the detection is tentative are marked.

$C_nH_mO$ bearing molecules are as abundant in the ROSINA mass spectra as pure hydrocarbons (Altwegg et al., 2017). These are mostly alcohols up to pentanol (5 carbon). In addition there are also aldehydes and acetone. Again there is a clear correlation between dust impacts and higher mass oxygen bearing hydrocarbons. The more dust is registered by COPS the more heavy molecules can be detected. Molecules with two oxygen atoms are on a 10% level comparable to the $C_nH_mO_1$ organics and are made up of acids and ethylene glycol. Benzoic acid is an aromatic ring molecule. CHN-compounds are at a similar level as the $CHO_2$-compounds. Amines are clearly identified as well as hydrogen cyanide and acetonitrile. The amides make up the group of CHNO, together with iso-cyanic acid and glycine, the only amino acid detected. CHS- and CHOS- compounds are quite abundant, probably because sulfur and nitrogen have similar chemistry (see above). More work is needed to distinguish between organics embedded in ice and therefore presenting a nucleus source or organics being sublimating from dust grains in the coma presenting a distributed source.

The detection of glycine (Altwegg et al., 2016) was not completely surprising as glycine was also detected in Stardust samples (Elsila et al., 2009). It was, however, unclear if glycine exists in a volatile phase. Glycine was first detected during a close flyby of the S/C at the nucleus. This facilitated the investigation of the link between glycine and grains. Indeed, glycine seems to behave differently with cometocentric distance than the overall density in the coma, which makes a distributed (grain) source likely. Together with glycine methylamine was also found, which opens a possible pathway for the formation of amino acid without liquid water. Methylamine is only present together with glycine. Formation of glycine without liquid water can be made by photochemistry of methylamine and $CO_2$ ice (Bossa et al., 2010). This is consistent with chemistry in interstellar icy dust mantles (Garrod, 2013) or ultraviolet irradiation of ice (Bossa et al., 2010) and subsequent accretion in the comet.

## 5.7 Unsaturated organics

Saturated complex hydrocarbons, which in an astrochemical sense are hydrocarbons containing more than 5 atoms, evade detection in the ISM, clouds or star forming regions, mostly because of their small dipole moments. In contrast, quite a large number of unsaturated hydrocarbons have been detected (Herbst & van Dishoeck, 2009). Among them are predominantly $C_xH$, x=2…8.

From the ROSINA measurements, it is evident that unsaturated molecules also play an important role in comets. Schuhmann et al., (2018) estimate that the total amount of saturated aliphatic and aromatic hydrocarbons is lower than the amount of the unsaturated species. The potential contribution of cycloalkanes is calculated to be almost negligible, which gives a ratio of saturated / unsaturated hydrocarbons = 44 / 56. The same trend is also seen for species containing oxygen and/or nitrogen. Unsaturated species can be the product of incomplete hydrogen addition or destruction of saturated species by UV / cosmic rays. The unsaturated hydrocarbons in the comet have generally more than one hydrogen, contrary to what has been observed in the ISM. It is difficult to say if this is due to a different composition / production mechanism or just a detection bias. Again, there is a clear correlation between the amount of detected molecules, their masses and the dust seen either by ROSINA-COPS and / or the star trackers of Rosetta. Heavier species appear only with dust and are mostly seen during the period of frequent outbursts of the comet near perihelion.

The COSIMA instrument on Rosetta analyzed cometary refractories. One of the important findings is that organics make up almost 50% of the refractories by weight (Bardyn et al., 2017). Another important finding is that the carbon in dust particles emitted by comet 67P (Fray et al., 2016) is bound in very large macromolecular compounds. This is similar to the insoluble organic matter (IOM) found in the carbonaceous chondrite meteorites, although the macromolecular compounds are less modified before and/or after being incorporated into the comet than the IOM found in meteorites. As ROSINA has seen spurious peaks up to m/z =172, the end of its mass range, it is very likely that we have to revise our notion that there are volatiles in a comet on one hand and refractories on the other hand. There is most likely a smooth transition from the very volatile species like $CH_4$ to insoluble macromolecules in the refractory part.

There were several attempts to classify comets according to their $C_2$- and $C_3$- abundances. Taxonomic classification of comets is based (in part) on their volatile composition by identifying a "typical" composition and then classifying the comets according to their enrichment/depletion. Systematic taxonomic classifications of comets have been done covering a number of radical species ($C_2$, $C_3$, CN, NH, $NH_2$, OH, O(1D)), and the dust continuum measured as Afρ (a quantity consisting of the dust albedo, the filling factor and the radius of the coma) describing the dust flux as these quantities are accessible to remote sensing. (Cochran et al. 2012, 130 comets; A'Hearn et al. 1995, 85 comets; Fink 2009, 50 comets; Langland-Shula & Smith 2011, 26 comets).

Despite decades of trying to clearly identify the parent molecules of these radicals, there are only a few established facts. Generally, NH and $NH_2$ radicals are attributed to $NH_3$ photodissociation. HCN is probably the main parent of CN, but not the only one, although no alternative has been identified so far. $C_2$ has been connected with parents such as $C_2H_2$ and $C_2H_6$ (Helbert et al. 2005) with mixed success. More recently, Weiler (2012) studied the chemistry of $C_2$ and $C_3$ in cometary comae and concluded that starting from $C_2H_2$, $C_2H_6$ and $C_3H_4$, the influence of ethane on the $C_2$ production has to be very minor. Looking at the data from Rosetta, we conclude that there are most probably many parents for $C_2$ as well as for $C_3$. $C_3H_4$ is not the most abundant $C_3$ component and is therefore probably not the only main contributor to $C_3$. There are many saturated species containing 2 or more C atoms. However, there are even more unsaturated species which can easily be photo destructed in the coma to yield either $C_2$ or $C_3$. E.g it is known that the photo-destruction rate of benzene is rather high (1.1 x $10^{-3}$ s, Crovisier et al., 1994) and leads via $C_6H_5$ to the $C_3$ radical. As the amount of heavier species (> $C_3$) is connected to dust, the classification according to $C_2$ and $C_3$ may be heavily biased by the amount of dust released by the comet. For 67P, it has been found that depending of the hemisphere (southern vs. northern), the comet can be classified as being depleted or normal (Le Roy et al., 2015). This is due to the different evolution of the surfaces. Whether a comet is classified in one category or another may depend on when exactly it was observed, the specific geometry of the comet (spin axis), and the evolution of its surfaces, and not due to a fundamental different composition.

Volatile organics represent in total mass about 1% of the volatiles. Organics in the refractories are almost 50% of the total mass of refractories. With the amount of water deduced from the Xe-isotopes in chapter 4.7 which we estimate to be about 1% of the terrestrial surface water, we get a very substantial amount of carbon delivered to the early Earth. Even if most organics did not survive the impact into the Earth atmosphere and to the surface, the amount of organcis still represents a significant part of today's total carbon inventory (Chyba et al., 1990).

## 5.8 Summary of molecular composition

Overall, cometary volatiles detected in 67P present a very rich chemistry, much more so than anticipated. Looking at table 4, 67P has more than doubled the number of parent species from 27 to 66. 39 of the 66 parent species have also been detected in clouds of low and high mass stars. For the others, detection may be difficult due to weak transition lines (e.g., for aliphatic saturated hydrocarbons or noble gases) or the abundance might be too small (e.g., for HBr). However, the many unidentified lines in high-resolution spectra from ALMA may contain more "cometary" molecules. On the other hand, the analysis of the ROSINA data is by far not finished and more detections are likely to come. For many of the species, there are no relative abundances available yet due to time-consuming calibration efforts of the instruments. Once this calibration is available, more meaningful comparison with other comets and the ISM will be possible.

So many molecules are common to comets and to the pre-stellar stages and available molecular abundances are very similar to those in the ISM. Figure 9 shows a comparison of data from 67P and from dark clouds, high and low mass protostellar clouds, all normalized to methanol. A perfect match would be on the diagonal line. In general and considering the uncertainties of observations, they all follow the same trend. There are a few outliers, namely the cooler, less dense outer part of the envelope around the low mass protostar IRAS 16293–2422 (16293c) and the dark cloud L134N. This might be due to regions where molecules may only partly be in the gas phase, especially as for the warmer part of the low mass protostar region IRAS 16293-2422 (16293w) the correlation with 67P is quite good. Surprisingly, also high mass protostar regions have similar abundances for complex organics as 67P. More data, especially on low mass star formation regions become now available. An interesting cross-check will be a comparison between the results from the PILS survey by Jorgensen et al. (2016;2018) and 67P in the near future.

These similarities suggest a common origin, as already deduced from the isotoplogues of water and other species. The presence of $O_2$, $S_2$, PO and others also clearly points to a pre-solar origin for these species at least.

# 6 Elemental abundances

Fig. 10, left panel shows a comparison of the main elemental abundances for dust and volatiles, extending the plot in Geiss et al. (1987) to 67P. Dust data used are from Bardyn et al. (2018). As the dust to ice ratio is still being debated for 67P, we show the values for a 1:1 and a 3:1 dust/ice ratio. For comparison, we also show values for the solar nebula, comet 1P/Halley with a dust/ice ratio of 1:2, for carbonaceous chondrites and the Earth (mantle and crust). Apart from 67P, 1P/Halley is the only comet where we have knowledge about dust and volatiles. It is clear from this plot that 1P and 67P are not fundamentally different, although they belong to different comet families. The differences can mostly be explained by the different dust/ice ratios assumed. Both comets have their full complement of C and O, but are deficient in N. This N-deficiency is still unclear and seems to be slightly more pronounced for 67P than for 1P.

Especially the $NH_3$ to water ratio is almost a factor 10 larger for comet 1P (1.5%, Meier et al. 1994) than for 67P. Dello Russo et al. (2016) have shown that comets coming close to the Sun generally have much higher $NH_3$ abundances. 1P was encountered by Giotto at 0.9 au outbound whereas the data for 67P are between 1.7 and 1.4 au inbound. As discussed previously, $NH_3$ seems to be, at least partly, in a more refractory phase, probably as ammonium salt (Altwegg et al., 2019) than as pure $NH_3$ in $H_2O$, which might even explain the apparent general N deficiency in comets. Assuming that the true $NH_3/H_2O$ abundance in 67P is close to what is seen in comets with small perihelion distances ($NH_3/H_2O$ ~3.5 %) and where therefore ammonium salts are efficiently released from dust (dello Russo et al., 2016), would bring the N-abundance of 67P and 1P almost to solar (see fig. 10, $N^*$).

Compared to meteorites, comets preserve much more of the volatile elements relative to silicon. This is not only due to the ices, which are missing in meteorites, but also due to the large amount of organics containing C and O in the dust (Bardyn et al., 2017). The comparison with meteorites clearly shows that the notion that comets represent the most pristine material in the solar system is correct not only for OCC's but also for JFC's.

Fig. 10, right panel shows elemental abundances in 67P relative to oxygen. Comets do not condense $H_2$ as they are not cold enough and not massive enough. Noble gases are deficient compared to solar, probably because they did not freeze out or have been lost in the Kuiper belt or the Centaur stage. The deficiency of 67P in some heavier elements like sulfur or chlorine relative to CI meteorites and solar is probably due to the fact that the cometary abundances given here represent only the ice of the comet, and not the dust. For a complete picture we have to wait for more results from the dust analyzer COSIMA.

# 7 From the ISM to comets

Cometary volatiles are probably the best accessible proxies to pre-solar material. Looking again at fig. 1 it is clear that cometary chemistry indeed reflects all of the stages from the ISM to life on Earth. Noble gas isotopes point to the nucleosynthesis at work feeding the ISM. The non-solar isotopic ratios, especially for xenon, are a clear indication of a poorly mixed cloud and proto-solar nebula. Traces of cloud chemistry are found in the sulfur species, especially the semi-volatile sulfur polymers, which have to be the result of photolysis or radiolysis of $H_2S$ on grain surfaces. $O_2$ is most likely the result of gas-grain chemistry in a relatively warm dark cloud. The large overlap in molecules seen in 67P and in clouds / star forming regions and their similar abundance ratios may point to a common origin. The high [$D_2O$/HDO] vs [HDO/$H_2O$] ratio can be explained by chemical models looking at the evolution from the pre-solar stage through the collapse of the disk.  The fact that this ratio is preserved in comets and the presence of a very volatile sulfur component $S_2$ imply that the ice never sublimated in the protoplanetary nebula. This as well as maybe rapid grain growth limits the chemical processes to which cometary ices have been exposed in the protoplanetary disk. The environment in the disk midplane where comets may have formed does not allow for much chemistry due to low temperatures, and low UV and cosmic ray fluxes although the latter is still being debated (Eistrup et al., 2016; 2018).

The many non-solar isotopic ratios detected in 67P's volatiles suggest a protoplanetary disk which is not homogeneized. The fluffy dust detected in 67P (Rotundi et al., 2015), the high porosity of dust grains found by the COSIMA instrument (Langevin et al., 2016) and the high porosity of the overall comet (Paetzold et al., 2015) do not allow for fast collisions during accretion, but rather for a gentle accretion. This gentle accretion is probably in line with the so-called pressure bumps seen by ALMA in the dust continuum in several protoplanetary disks (e.g. Van der Marel et al., 2013).

In the Stardust samples from Wild 2, high temperature refractories (CAI's) have been detected, which indicate radial mixing in the disk. CAI's are thought to be formed in the inner solar system, well inside the water snow line.  In 67P, there is only one tentative detection of such a grain (Paquette et al., 2016). The large variation of D/H found in comets calls into question the notion that OCC's and JFC's have been formed at different locations in the protoplanetary nebula. The two families experienced different dynamical histories. It would be interesting to know if comets like Hartley 2, with a terrestrial D/H, contain a large abundance of CAI's or if Wild 2 has a low D/H in water. Unfortunately, this information must wait for encounters with these comets.

The mean D/H for comets of ~ $3.6 \times 10^{-4}$ - although the sample may still be rather small – tells us that the Earth did not get its bulk surface water from comets. However, from the isotopic signature of xenon in our Earth atmosphere and the signature of xenon in 67P, we can deduce the amount of water (~1% of terrestrial surface water, Marti et al., 2017) and therefore of carbon / organics delivered by comets. While the amount of water is rather insignificant and does not dominate the terrestrial D/H ratio, the amount of carbon is an important contribution to the Earth's surface carbon inventory. Together with the rich organic chemistry observed in comets, this may make the notion that comet impacts sparked life on Earth more probable.

As most of these molecules are probably pre-stellar, similar conditions as on Earth with impacting comets could exist elsewhere. However, the detection of many molecules in a comet, which were before classified as biomarkers, makes the detection of biological activity elsewhere much more complicated. Molecules like e.g. $CH_3Cl$, if detected in a planetary atmosphere, may have an abiotic origin. Combinations of molecules like abundant $O_2$ combined with $CH_4$ were thought to be indicative of ongoing biology, but of course this is not so in comets. To rule out false detection of biological processes will therefore be a very major challenge when looking for life on exoplanets.

# 8 Summary points

- Comets are the most pristine bodies found in our solar system as demonstrated by their almost solar abundance of the light elements C, N and O.
- Comets contain a very rich complement of organics with abundances similar to presolar material found in clouds and star forming regions.
- Many isotopic ratios, e.g. Si, Xe and S in comets are non-solar, pointing to a not well-mixed protoplanetary disk.
- The wide range of D/H values found in cometary water independent of the comet family point to a formation of comets over a large radial distance in the early solar disk.
- The high [$D_2O$/HDO] / [HDO/$H_2O$] ratio found in comet 67P are indicative of presolar, never sublimated cometary water.
- The high abundance of $O_2$ is indicative of gas-grain chemistry.
- The presence of sulfur polymers $S_2$-$S_4$ point to grain surface chemistry through photolysis / radiolysis.
- Three noble gases, namely Ar, Kr and Xe have been detected in the coma of 67P, whereby Ar seems to be depleted relative to the other two, probably because JFC's have lost highly volatiles on their way inwards from the Kuiper belt / Scattered disk.
- The xenon isotopic ratios measured in 67P can explain the long standing question about the origin of the terrestrial atmospheric xenon. Together with the noble gases the amount of organics delivered by comets may be highly significant.
- 

# 9 Future issues for cometary science

- The upcoming JWST mission will be able to directly observe ice in the interstellar medium as well as in comets. This will help to better understand the ongoing chemistry in clouds and star forming regions. Together with ALMA observations and future cometary missions, the different stages from the ISM to planets and maybe life can be followed studying and comparing their chemical compositions.
- In order to solve the problem of heterogeneity between comets, the next comet mission should target a comet as different from 67P as possible, e.g. Hartley 2 where we know that D/H is low.
- Main belt comets in the asteroid belt may present an intermediate population between comets and asteroids. Exploring such a body from close up, especially looking at deuterated water, is an important step towards understanding the formation of these icy bodies.
- Exploring other icy bodies, especially moons of Jupiter and Saturn with in situ instrumentation could also give clues on transport of matter and chemistry in the protoplanetary disk.

- Future cometary missions should carry /consist of a larger lander, being able to look at the interior of the comet to at least > 1m, hopefully much deeper.
- Sample return of refractories opens the possibility to study mineralogy of the organics, which was largely missed by Stardust due to the high velocity impacts of dust grains.
- Cryogenic sample return below 60K (sublimation of $CO_2$) with a sample from the deep interior has to be the final goal of cometary science.

## DISCLOSURE STATEMENT

The authors are not aware of any affiliations, memberships, funding, or financial holdings that might be perceived as affecting the objectivity of this review.

## Acknowledgment

The authors would like to thank the following institutions and agencies, which supported this work: Work at University of Bern was funded by the State of Bern, the Swiss National Science Foundation and by the European Space Agency PRODEX Program. ROSINA research at Southwest Research Institute is funded by NASA through the Jet Propulsion Laboratory. All ROSINA data are the work of the international ROSINA team (scientists, engineers and technicians from Switzerland, France, Germany, Belgium and the US) over the past 25 years, which we herewith gratefully acknowledge. Rosetta is an ESA mission with contributions from its member states and NASA. We gratefully acknowledge the many useful inputs from the editor and referee, EF van Dishoeck, which helped to significantly improve the manuscript.

*Table 1: $^{16}O/^{18}O$ in cometary water*

| $^{16}O/^{18}O$ | Comet | Reference |
|---|---|---|
| 518 ± 45 | 1P/Halley | Balsiger et al. (1995) |
| 470 ± 40 | | Eberhardt et al. (1995) |
| 530 ± 60 | 4 comets | Biver et al. (2007) |
| 425 ± 55 | C/2002 T7 | Hutsemékers et al. (2008) |
| 300 ± 150 | C/2012 F6 | Decock et al. (2014) |
| 523 ± 32 | C/2009 P1 | Bockelée-Morvan et al. (2012) |

Table 2: Overview of measured isotopic ratios in comets. For references see specific subchapters.

| **Isotopic ratio**<br>Solar / standard | **Species /**<br>ratio in 67P | | | **Other comets (average)** | **Minor relative to major isotope relative to Sun** |
|---|---|---|---|---|---|
| **D/H** 1.5 $10^{-5}$ VSMOV | **$H_2O$** (5.3±0.7) x $10^{-4}$ | **$NH_2D$** ~ 0.1 % | **HDS** ~ 0.06 % | **$H_2O$** 3.4 x $10^{-4}$ | enriched |
| | **$D_2O$/HDO** ~ 1 % | | | | enriched |
| **$^{12}C/^{13}C$** 89-98 | **$CO_2$** 84±4 | **CO** 86±9 | **$C_2H_5$** 84±12 | **$C_2$, CN, HCN** 95 | Slightly enriched in $CO_2$ |
| **$^{32}S/^{34}S$** 22.65 | **$H_2S$** 23.8 ± 0.5 | **$SO_2$** 24.3±1.2 | **$CS_2$** 25.6±0.6 | **CS, S+, H2S** 20.5 | depleted |
| **$^{32}S/^{33}S$** 126.95 | **$H_2S$** 183±8.1 | **$SO_2$** 146±9 | **$CS_2$** 175±13 | | depleted |
| **$^{29}Si$ /$^{28}Si$**5.08% | **Dust** (4.3±0.5)% | | | | depleted |
| **$^{30}Si/^{28}Si$**3.25% | **Dust** (2.6±0.6)% | | | | depleted |
| **$^{14}N$ / $^{15}N$** 440 | **$NH_3$** 118±25 | **NO** ~120±25 | **$N_2$** ~ 130±30 | **HCN, CN, NH2** 140 | enriched |
| **$^{16}O/^{18}O$** 530 | **$H_2O$** 445 ± 35 | **$CO_2$** 494 ± 8 | | **$H_2O$** 461 | enriched /normal |
| **$^{35}Cl/^{37}Cl$** 3.127 | **HCl** 3.4 ± 0.2 | | | | normal |
| **$^{79}Br$ / $^{81}Br$** 1.077 | **HBr** 1.05 ± 0.08 | | | | normal |
| **$^{36}Ar/^{38}Ar$** 5.3 | 5.4 ± 1.4 | | | | normal |
| **$^{80}Kr/^{84}Kr$**<br>**$^{82}Kr/^{84}Kr$**<br>**$^{83}Kr/^{84}Kr$**<br>**$^{86}Kr$ / $^{84}Kr$** | 0.036 ± 0.022<br>0.206± 0.012<br>0.187 ± 0.011<br>0.287 ± 0.014 | | | | normal except $^{83}Kr$ |
| **$^{128}Xe/^{132}Xe$**<br>**$^{129}Xe/^{132}Xe$**<br>**$^{130}Xe/^{132}Xe$**<br>**$^{131}Xe/^{132}Xe$**<br>**$^{134}Xe/^{132}Xe$**<br>**$^{136}Xe/^{132}Xe$** | 0.08 ± 0.023<br>1.463 ± 0.136<br>0.199 ± 0.053<br>0.888 ± 0.101<br>0.212 ± 0.046<br>0.113 ± 0.047 | | | | enriched in $^{129}Xe$, depleted in $^{134,136}Xe$ |

Table 3: relative abundances of halogens vs. oxygen in 67P, the Sun, chondrites and Earth.

| **Halogenes** | **67P (x10$^{-4}$)[a]** | **Sun(x10$^{-4}$)[b]** | **Chondrites(x10$^{-4}$)[c]** | **Earth(x10$^{-4}$)[c]** |
|---|---|---|---|---|
| F/O | 0.2 – 4 | 0.51 ± 0.12 | 1.1 ± 0.2 | 0.16 ± 0.02 |
| Cl/O | 2 - 5 | 3.29 ± 0.77 | 6.8 ± 1.2 | 0.31 ± 0.16 |
| Br/O | 0.01-0.07 | 0.0068 ± 0.0002 | 0.014 ± 0.003 | 0.012 ± 0.04 |

[a] Coma of 67P from Dhooghe et al. (2017)

[b] Early Sun from Lodders (2010).

[c] Earth derived from Allègre, Manhès & Lewin (2001).

Table 4. Parent species detected in the ROSINA DFMS mass spectra and comparison with the ISM

a) **Inorganics**

| Chemical formula | Name | ISM[1] | m/z | Chemical formula | Name | ISM | m/z |
|---|---|---|---|---|---|---|---|
| *$NH_3$* | *ammonia* | y | 16 | Ar | argon | n | 36 |
| *$CH_4$* | *methane* | y | 17 | *$CO_2$* | *carbon dioxide* | y | 44 |
| *$H_2O$* | *water* | y | 18 | PO | phosphorous monoxide | y | 47 |
| HF | hydrogen fluoride | y | 20 | *SO* | *sulfur monoxide* | y | 48 |
| *CO* | *carbonmonoxide* | y | 28 | *$S_2$* | *sulfur dimer* | n | 64 |
| $N_2$ | nitrogen | n | 28 | *$SO_2$* | *sulfur dioxide* | y | 64 |
| NO | nitrous monoxide | y | 30 | $CS_2$ | *carbonyl sulfide* | y* | 76 |
| $O_2$ | oxygen | y | 32 | HBr | hydrogen bromide | n | 80 |
| *$H_2S$* | *Hydrogen sulfide* | y | 34 | Kr | krypton | n | 84 |
| $H_2O_2$ | Hydrogen peroxyde | n | 34 | $S_3$, $S_4$ | Sulfur polymers | n | 96, 128 |
| HCl | hydrogen chloride | y[1] | 36 | Xe | xenon | n | 132 |

* only the radical CS is observed

b) **Organics**

| Chemical formula | Name | ISM[1] | m/z | Chemical formula | Name | Found in ISM | m/z |
|---|---|---|---|---|---|---|---|
| **CH -compounds** | | | | **CHO- compounds** | | | |
| *$CH_4$* | *Methane* | y | 16 | *$CH_2O$* | *Formaldehyde* | y | 30 |
| *$C_2H_2$* | *Ethylene* | y | 26 | *$CH_4O$* | *Methanol* | y | 32 |
| *$C_2H_6$* | *Ethane* | n | 30 | *$C_2H_4O$* | *Acetaldehyde* | y | 44 |
| *$C_3H_8$* | *Propane* | n | 44 | *$C_2H_6O$* | *Ethanol* | y | 46 |
| $C_4H_{10}$ | Butane | n | 58 | *$C_3H_6O$* | *Acetone* | y | 58 |
| $C_5H_{12}$ | Pentane | n | 72 | $C_3H_8O$ | Propanol | n | 60 |
| $C_6H_6$ | Benzene | y | 78 | $C_4H_{10}O$ | Butanol | n | 74 |
| $C_7H_8$ | Toluene | n | 92 | $C_5H_{12}O$ | Pentanol | n | 88 |
| $C_7H_{16}$ | Heptane | n | 98 | **$CHO_2$-compounds** | | | |
| $C_8H_{10}$ | Xylene[2] | n | 106 | *$CH_2O_2$* | *Formic acid* | y | 46 |
| $C_8H_{16}$ | Octane [2] | n | 112 | *$C_2H_4O_2$* | Acetic acid, *Glycolaldehyde* | y<br>y | 60 |
| $C_{10}H_8$ | Naphthalene [2] | n | 128 | *$CH_2(OH)CH_2(OH)$* | *Ethylene glycol* | y | 62 |
| **CHN-compounds** | | | | $C_7H_6O_2$ [2] | Benzoic acid | n | 122 |
| | | | | **CHON-compounds[3]** | | | |
| *HCN /HNC* | *Hydrogencyanide* | y | 27 | CHNO | Isocyanic acid | y | 43 |
| $CH_5N$ | Methylamine | y | 31 | $CH_3NO$ | Formamide | y | 45 |
| *$CH_3CN$* | *Acetonitrile* | y | 41 | $CH_3CONH_2$ | Acetamide | y | 59 |
| C2H7N | Ethylamine | n | 45 | $C_2H_5NO_2$ | Glycine | n | 75 |
| *HC3N* | *Cyanoacetylene* | y | 51 | **CHS-compounds** | | | |
| $C_3H_9N$ | Propylamine | n | 59 | $CH_2S$ | Thioformaldehyde | y | 46 |
| **CHOS-compounds** | | | | $CH_4S$ | Methanethiol | y | 48 |
| *COS* | *Carbonylsulfide* | y | 60 | $C_2H_6S$ | Ethanethiol [4] | y | 62 |
| $CH_4OS$ | [5] | y | 64 | $(CH_3)S(CH_3)$ | Dimethylsulfide [4] | y | 62 |
| $C_2H_6OS$ | [5] | n | 78 | **$CHS_2$-compounds** | | | |
| $C_3H_6OS$ | [5] | n | 90 | $CH_4S_2$ | Methyl hydrogen disulfide | n | 80 |

Species in italics have been detected also in other comets

[1] Müller et al., 2001; Müller et al., 2005; https://www.astro.uni-koeln.de/cdms/catalog (07/2018)).

[2]Tentative detection

[3] hydrocarbons containing oxygen and nitrogen, not to be confused with CHON grains seen in comet 1P/Halley, consisting mostly of C, O, N and H atoms

[4] Isomers

[5] Unclear which isomer is responsible as there are several choices

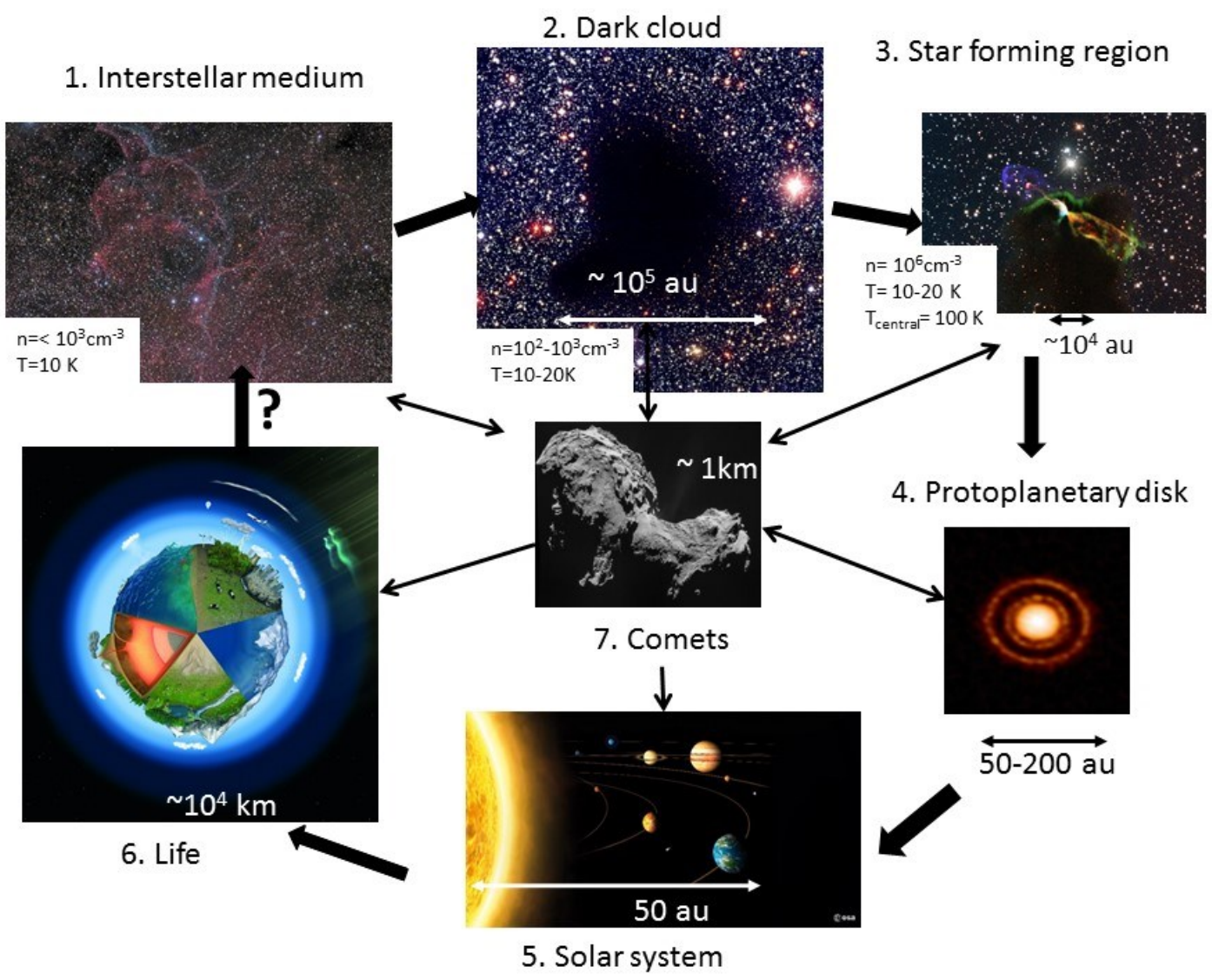

Figure 1: Different stages of the formation of planetary systems. Starting in the diffuse interstellar medium with material from dying stars and supernovae (stage 1) the atoms form molecules which undergo chemistry in giant molecular clouds containing many times the mass of the Sun. Part of the cloud can centrally condense to higher densities forming a dense core, sometimes also called a pre-stellar core (stage 2). This core can collapse to form a protostar with a young disk and an outflow that clears the surrounding cloud (stage 3). When the young star emerges from its cocoon, it is surrounded by a protoplanetary disk, in which planet formation can occur. Its temperature and density decrease with radial distance from the star (stage 4).The protoplanetary disk that formed our own solar system is also called the protosolar nebula. After dissipation of the remaining dust and gas a star with planets evolves over billion years (stage 5). Eventually, life may start on a planet (stage 6). At the end of the Sun's lifetime, its material will be fed back into the interstellar medium. Comets contain material from all stages of this evolution (image 7).
Scales given (within an order of magnitude) are typical for these different stages. Credit for pictures 1-4: European Southern Observatory; for 5-7: European Space Agency

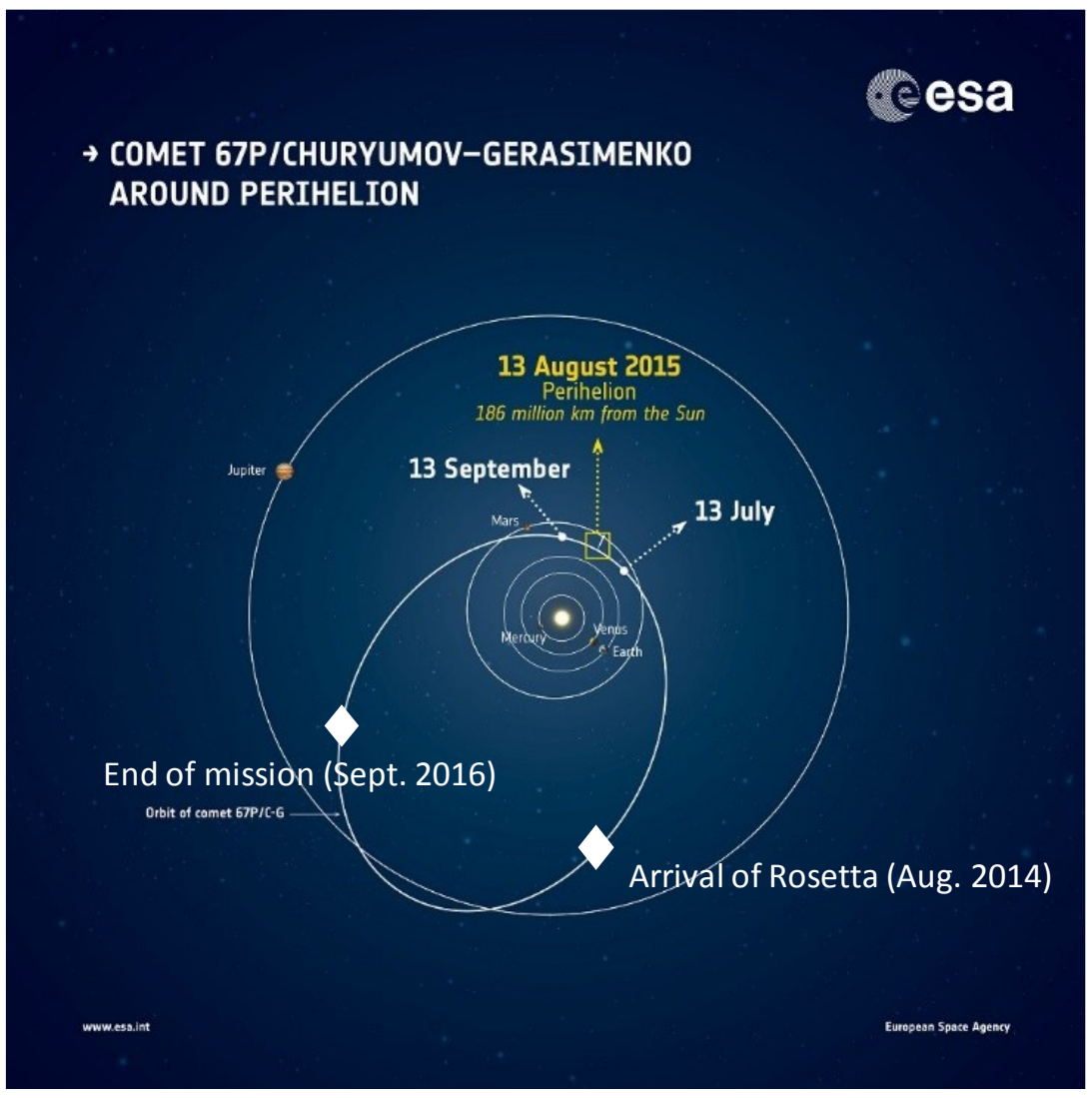

Fig. 2: Orbit of 67P/Churyumov-Gerasimenko, Credit: ESA

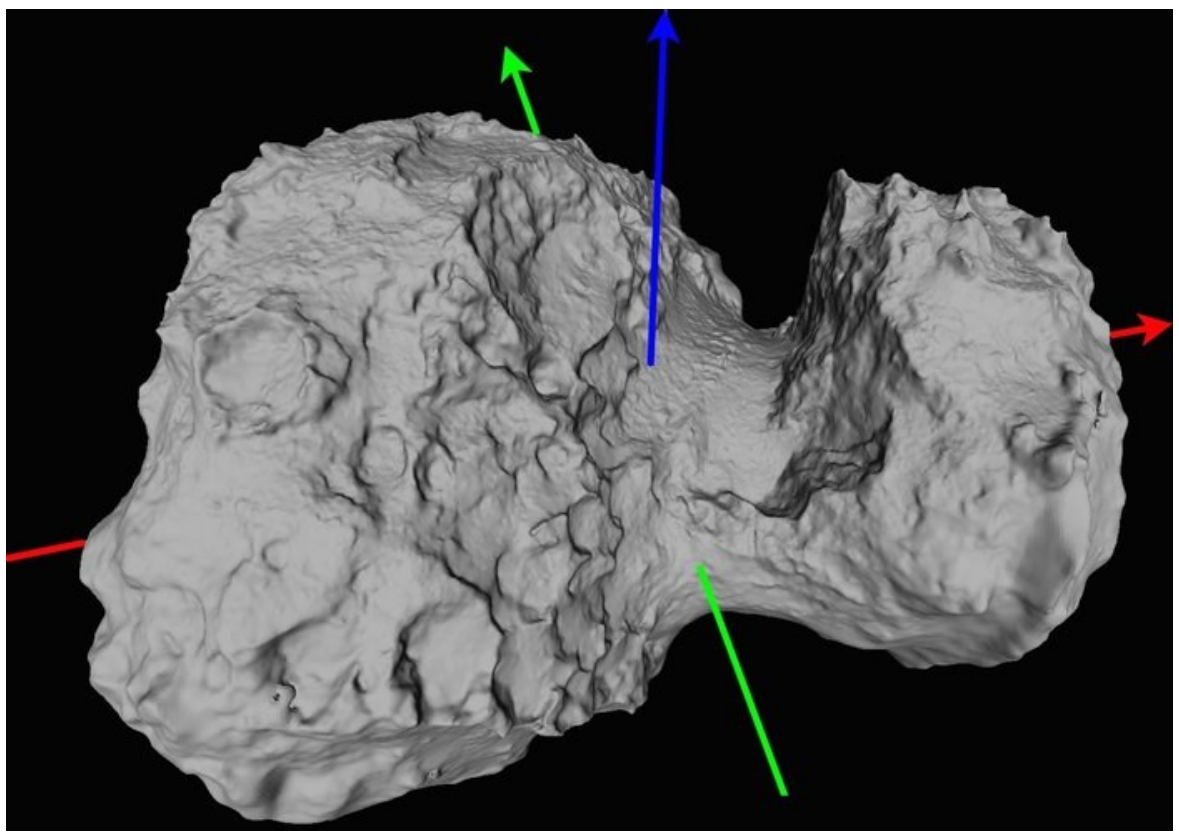

Fig. 3: 67P/Churyumov-Gerasimenko. The blue arrow indicates Comet 67P/Churyumov–Gerasimenko's rotation axis, and the red and green arrows display its equatorial x- and y-axes, respectively. Its spin axis is tilted by 52º.

Credits: ESA/Rosetta/MPS for OSIRIS Team MPS/UPD/LAM/IAA/SSO/INTA/UPM/ DASP/IDA

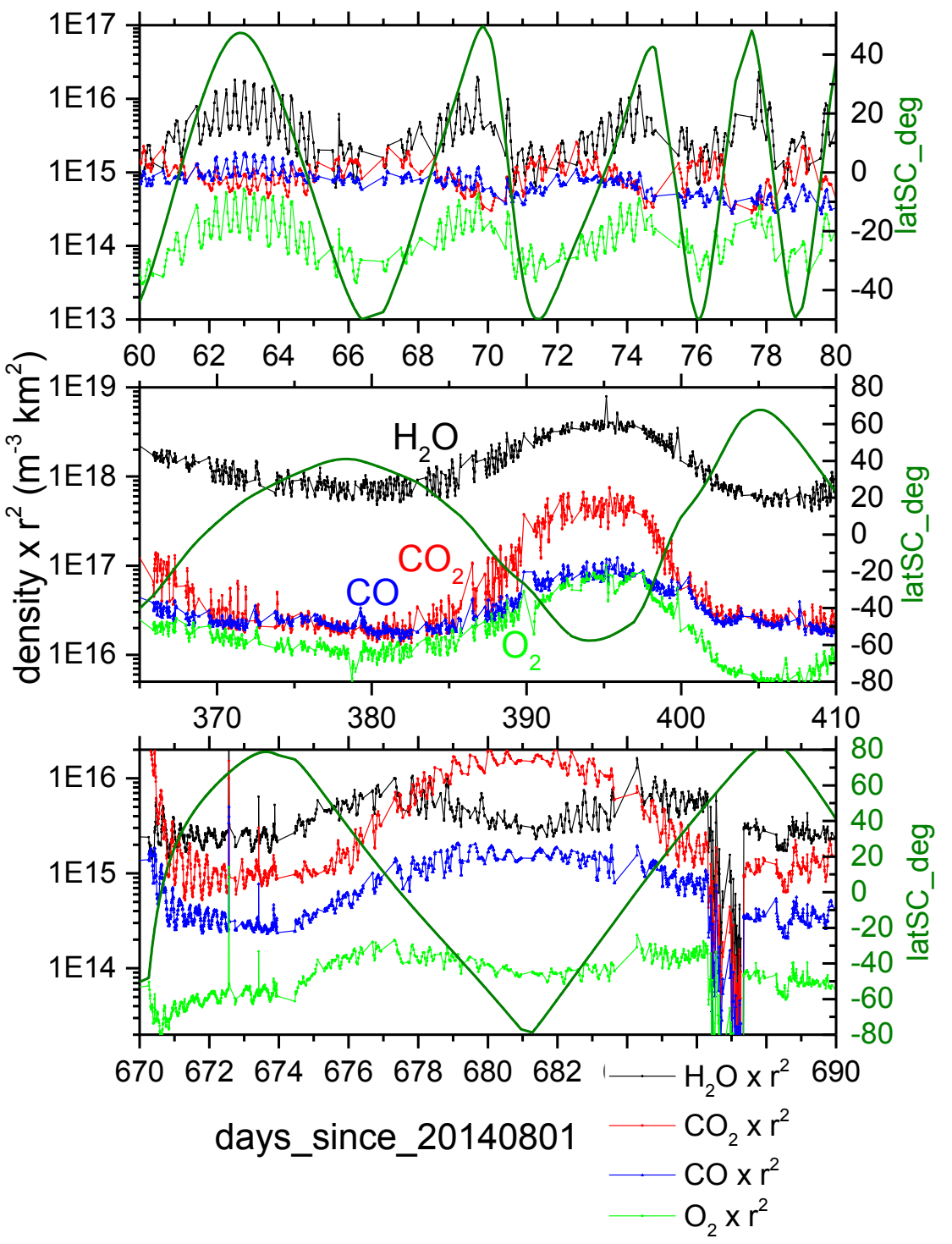

Fig. 4: Comparison of CO, $CO_2$ and $H_2O$ local densities multiplied by $r^2$ for three different periods: a) > 3 au, inbound; b) around the highest activity of the comet at 1.25 au and c) > 3 au, outbound. Also plotted are the sub-S/C latitudes (dark green line) (>0: north; <0: south). It is evident that relative abundances vary with time and subspacecraft latitude (season). Evident is the fact that $O_2$ follows water for most of the orbit.

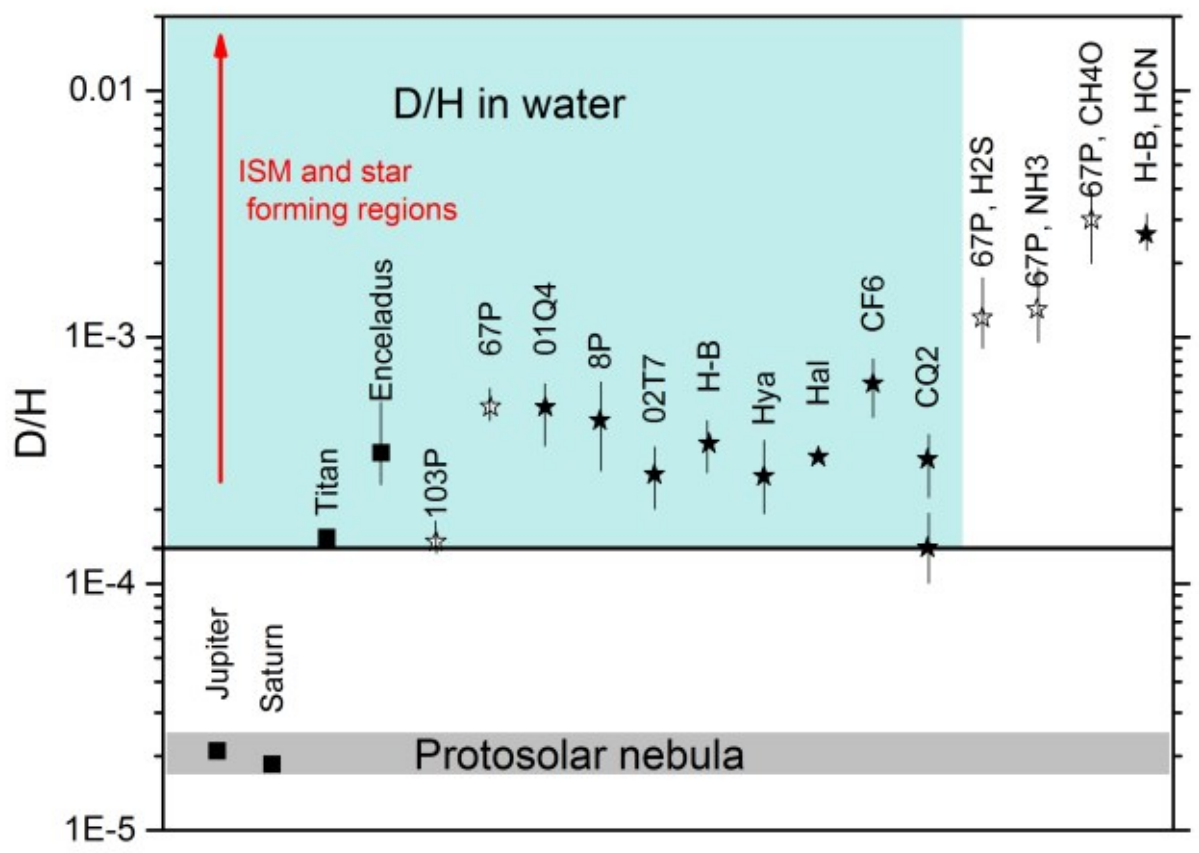

Fig. 5 Overview of measured D/H in water and other cometary species. For comparison, we also show the giant planets, Titan and Enceladus and the protosolar nebula. Filled stars denote OCC's. The black line denotes the Earth value. 01Q4: C/2001 Q4 (NEAT) ;8p: 8P/Tuttle; 02T7: C/2002 T7 (LINEAR) ;H-B: Hale-Bopp; Hya: Hyakutake; CF6: C/2012 F6 (Lemmon); CQ2: C/2014 Q2 (Lovejoy). 67P: water and $H_2S$: Altwegg et al. (2017); $NH_3$: Wampfler et al. (2018). CQ2: Paganini et al. (2017); Biver et al. (2016); CF6: Biver et al. (2016). All other values from Bockelée-Morvan (2015) and references therein.

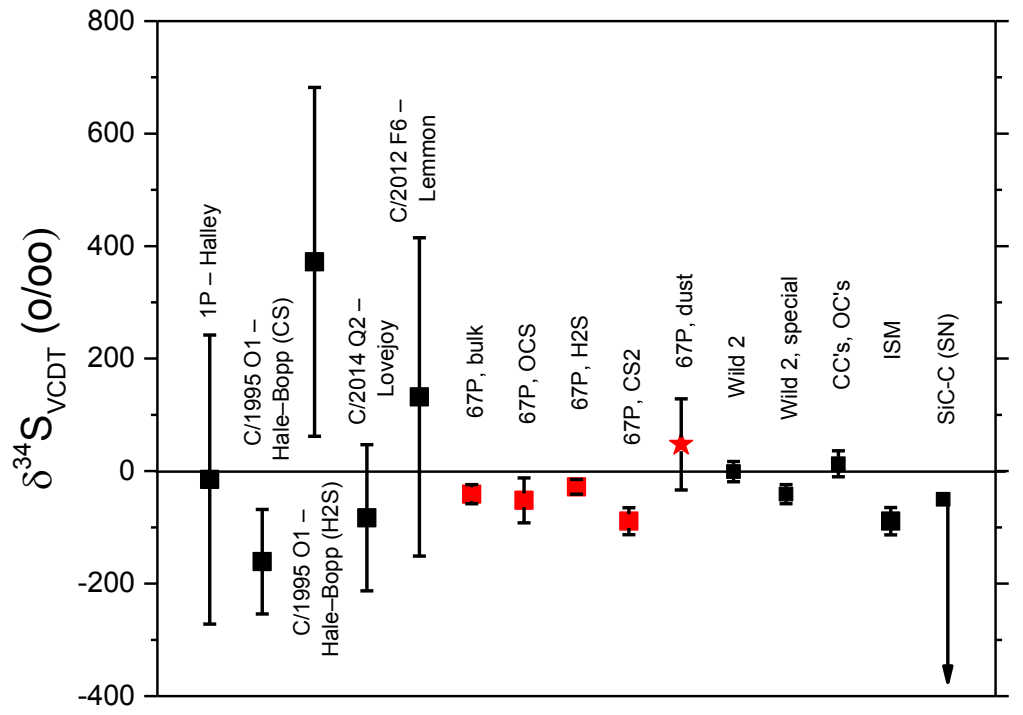

Fig. 6 Sulfur-isotopic compositions, given as permil deviation from the VCDT standard, of bulk carbonaceous and ordinary chondrites (CCs/OCs), residues in impact craters on Al foils from NASA's Stardust mission to comet 81P/Wild 2, different molecules (as well as their weighted mean and dust) from comet 67P/CG, presolar SiC Type C grains from SN explosions. Data sources: chondrites: (Bullock et al. 2010; Gao and Thiemens 1993a,b); 81P/Wild 2: (Heck et al. 2012); 67P/CG: (Calmonte et al. 2017); 67P/CG dust : (Paquette et al., 2017) ; presolar SiC: (Gyngard et al. 2010; Hoppe et al. 2012; Liu et al. 2016; Xu et al. 2015); ISM (Mauersberger et al. 2004).Results from 67P's volatiles are clearly non-compatible with the meteoritic standard, but agree well with the local ISM value derived by Mauersberger et al. 2004).

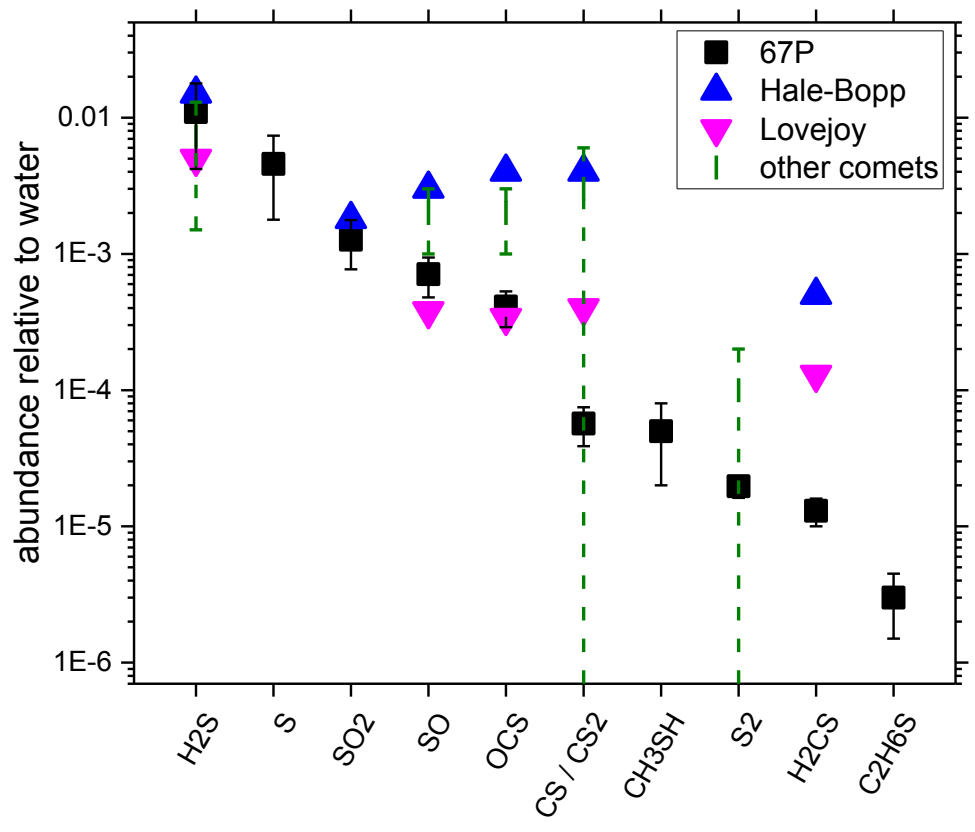

Figure 7: Sulphur bearing species detected in comets, including 67P (from Calmonte et al., 2016 and references therein).

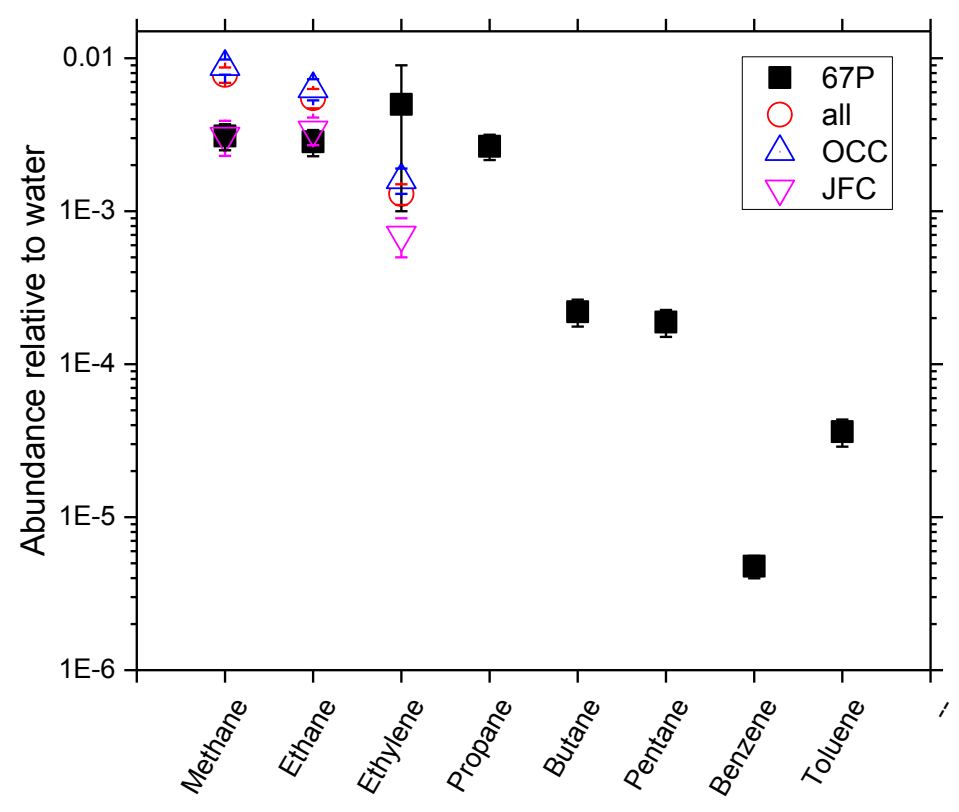

Figure 8: Saturated aliphatic and aromatic hydrocarbons detected in the coma of 67P (Schuhmann et al., 2018) close to perihelion.

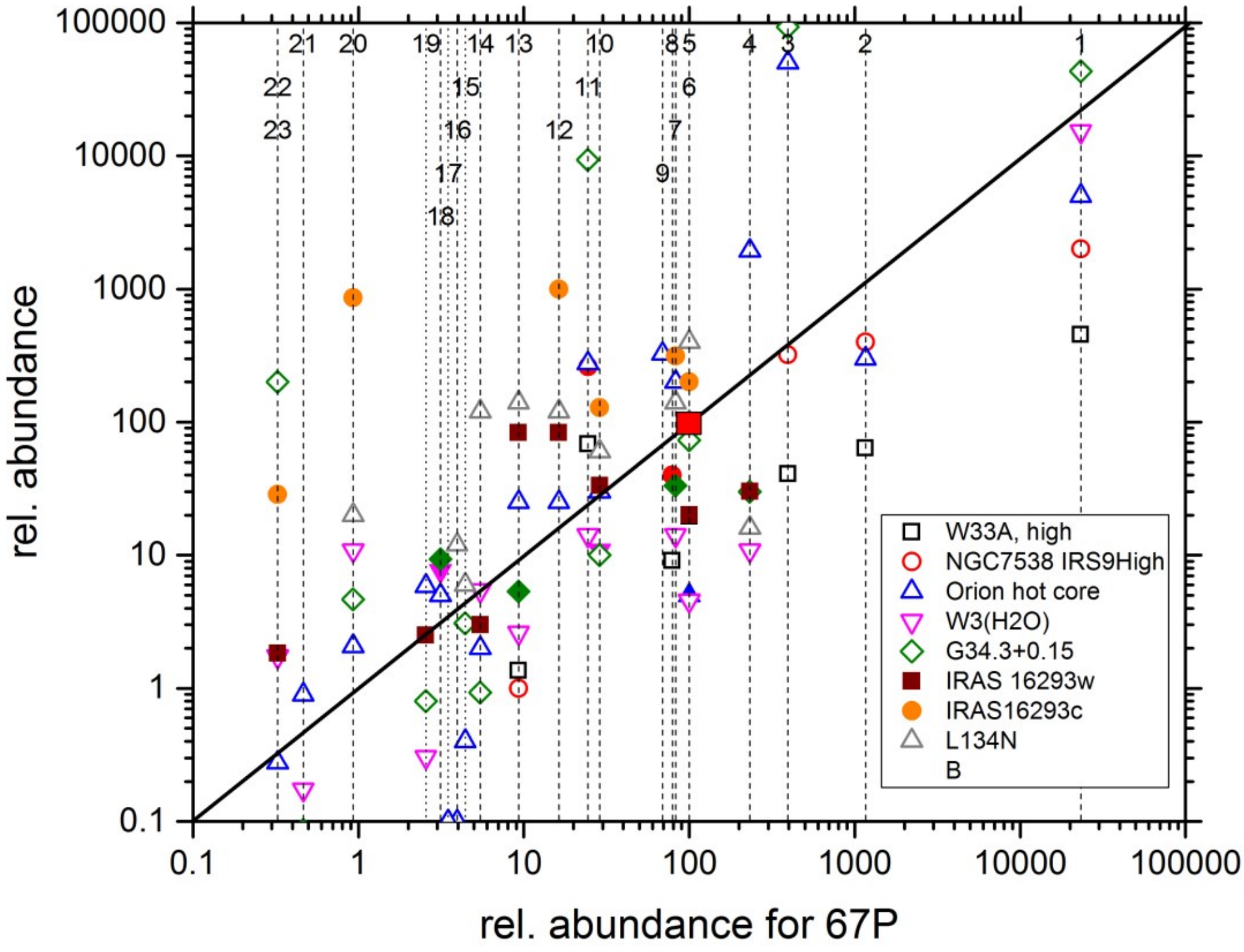

Fig. 9: Comparison of molecular abundances relative to methanol (red square) in clouds of high and low mass protostars with 67P. 1: $H_2O$; 2: $CO_2$; 3: CO; 4: $H_2S$; 5: $H_2CO$; 6: $CH_3OH$; 7: HCN; 8: $CH_4$; 9: $C_2H_2$; 10: $SO_2$; 11: $NH_3$; 12: SO; 13: OCS; 14: HNCO; 15: HCOOH; 16: $CH_3CHO$; 17: $NH_2CHO$; 18: $HCOOCH_3$; 19: $CH_3CN$; 20: CS/$CS_2$*; 21: $HC_3N$; 22: $CH_3SH$; 23: $H_2CS$. Data: W33A from Gibb et al 2000b, Keane et al 2000, NGC7538:IRS9 from Whittet et al 1996, Schutte 1999, Ehrenfreund & Schutte 2000, Keane et al 2000; Orion hot core from van Dishoeck & Blake (1998); Irvine et al (1999), Sutton et al. (1995); L134N from Ohishi et al. (1992), updated using values from (2000) for L134N pos. C; L1157 from Bachiller & Pérez Gutiérrez (1997) at position B2 assuming CO/H2 = $10^{-4}$. W3 from Helmich et al. (1996); Helmich&van Dishoeck (1997) ; Hermsen et al. (1988); G34.3+0.15 from Macdonald et al.(1996); Millar et al. (1997); Hatchell et al. (1998a) ; Hatchell et al. (1998b); Heaton et al. (1989); Hatchell et al. (1998c); Bockelee-Morvan et al., 2000. IRAS 16293w: Abundance in warm and dense inner part of the envelope 150 AU in radius around IRAS 16293–2422 from Schöier et al., 2002. IRAS16293c: Abundance in cooler, less dense outer part of the envelope around IRAS 16293–2422 from Schöier et al., 2002. Data for 67P from Le Roy et al., 2015; Calmonte et al., 2017; Altwegg et al., 2017 ; Schuhmann et al., 2018.

*Cometary data are for $CS_2$ while the other data are for the radical CS.

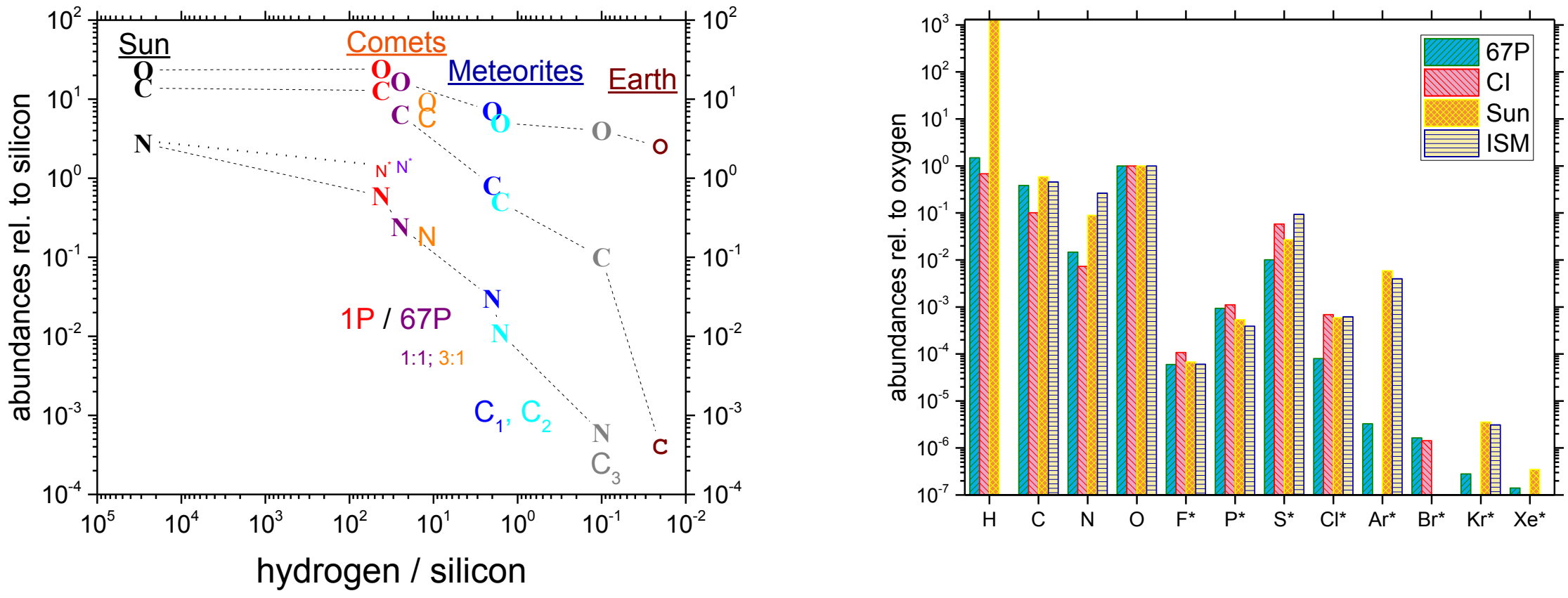

Fig 10: Left: Comparison of elemental abundances (after Geiss et al., 1987) for the Sun, comets (dust and ice), carbonaceous chondrites and the Earth (mantle and crust). For 67P, values are given for the case with dust/ice =1 and dust/ice =3. Dust abundances for 67P are from Bardyn et al. (2018). The values for 1P/Halley are for dust/ice=1:2. N* denotes nitrogen abundance if we assume a $NH_3/H_2O$ abundance of ~3.5 % as seen in comets with a very small perihelion distance (dello Russo et al., 2016). Values for the solar nebula, 1P/Halley, carbonaceous chondrites and the Earth are from Geiss et al. (1987) and references therein. Right: comparison of all volatile elements in 67P to CI meteorites (Palme et al., 2014), solar photosphere (Palme et al., 2014) and towards ζ Ophiuchus (Palme et al., 2014). Elements with asterisk contain only the volatile part for 67P as the refractory part is still unknown.